\documentclass[%
pra, twocolumn,
superscriptaddress,
groupedaddress,
amsmath,amssymb,
aps,
floatfix,
]{revtex4-2}

\usepackage{graphicx}
\usepackage{dcolumn}
\usepackage{bm}
\usepackage{graphicx}
\usepackage{hyperref}
\hypersetup{colorlinks=true, linkcolor=blue, citecolor=blue, urlcolor=blue}

\usepackage{amsmath}
\usepackage{subcaption}
\usepackage{tikz}
\usepackage{graphicx}
\usepackage{subcaption}
\DeclareCaptionJustification{revtex}{\flushing}
\usepackage{physics}

\newlength{\availW}
\newlength{\hL}
\newlength{\hR}
\newlength{\wL}
\newlength{\wR}
\newsavebox{\lbox}
\newsavebox{\rbox}

\tikzset{
	gate/.style={draw, thick, minimum width=2.2cm, minimum height=3cm, fill=white},
	line/.style={thick},
	every node/.style={font=\normalsize}
}

\begin{document}
	
	\preprint{APS/123-QED}
	
	\title{Modular linear-optics collision model for quantum readout}
	
	\author{Tanay Shardul Gangan}
	
	\affiliation{Department of Physics, Indian Institute of Technology, Hyderabad 502284, India}
	\author{Archak Purkayastha}%
	\email{archak.p@phy.iith.ac.in}
	\affiliation{Department of Physics, Indian Institute of Technology, Hyderabad 502284, India}
	\date{\today}
	
	\begin{abstract}
	Homodyne and heterodyne measurements of propagating fields are commonly described by stochastic master equations, whose derivation relies on stochastic calculus and must be adapted to each detection chain. We develop a modular linear-optics collision model in which each element of the chain is a gate unitary and each photodetector is a click operator. Composing these and tracing out ancillas in known states gives effective Kraus operators. The model avoids stochastic calculus, keeps individual photocounts, which matter for weak local oscillators and short integration times, and reproduces the standard master equation and measurement operators in the continuous-time limit. The Kraus operators of a simpler readout scheme serve as a module for a more complex one, obtained by adding or replacing gates. We show this with two examples. First, we derive a homodyne module and use it to build a heterodyne module. Applied to dispersive qubit readout, it reveals a fundamental assignment error that depends on the initial qubit state and persists at any finite integration time even in ideal readout, i.e., without qubit decoherence or detection imperfections. Since the photocounts remain discrete, the quadrature statistics also cross over from Skellam to Gaussian with increasing local-oscillator strength, as expected. Second, we add phase-preserving gain by replacing only the collision gate. The new three-mode gate combines beam-splitter coupling and two-mode squeezing, as in superconducting parametric couplers. The gain adds only the minimum noise allowed by quantum mechanics but also acts back on the readout mode. This back-action, which a description of the gain as added noise alone misses, is needed for the correct signal-to-noise ratio. Our framework thus provides a microscopic, practical, and modular route to modeling, and ultimately optimizing, complex quantum readout chains.
	\end{abstract}
	
	\maketitle
	
	
	\section{Introduction}
	\label{sec:intro}
	
	Continuous measurement of propagating electromagnetic fields is a central tool of quantum optics and quantum information processing~\cite{WisemanMilburn2009,JacobsSteck2006}. A prominent example is the dispersive readout of superconducting qubits, in which the qubit state is inferred from the microwave field that leaks out of a coupled readout resonator~\cite{Blais2004,Blais2021,Krantz2019}. Before this field is detected, it passes through a chain of linear elements, such as beam splitters, mixers driven by a local oscillator, and phase-preserving parametric amplifiers~\cite{Clerk2010,RoyDevoret2016,Macklin2015}. Each element modifies both the signal and the noise, so the attainable readout fidelity is set by the entire chain rather than by the final detection event alone~\cite{Walter2017,Heinsoo2018}.
	
	The conditional dynamics of a system monitored through such a chain is commonly described by a stochastic master equation (SME) within quantum trajectory theory~\cite{Dalibard1992,Gardiner1992,Carmichael1993,WisemanMilburn1993}. Homodyne and heterodyne detection correspond to particular diffusive unravelings of the unconditional master equation~\cite{WisemanDiosi2001,ChiaWiseman2011,WisemanMilburn2009}, and in circuit quantum electrodynamics, SMEs describe the dephasing and quantum jumps induced by dispersive measurement as well as the individual quantum trajectories observed in experiments~\cite{Gambetta2008,Murch2013,Hatridge2013}. For a realistic detection chain, however, each element has to be translated into effective parameters of the SME, such as a detection efficiency, the choice of unraveling, added noise, or additional drives. This translation is usually worked out case by case, and the resulting equation no longer shows how the individual elements contribute to the measured signal. Moreover, SMEs are stochastic differential equations whose derivation and manipulation rely on classical or quantum stochastic calculus~\cite{HudsonParthasarathy1984,GardinerCollett1985,GardinerZoller2004}, a technical barrier that has motivated several pedagogical treatments of continuous measurement~\cite{Brun2002,JacobsSteck2006,GenoniLamiSerafini2016,Gross2018}. Finally, the diffusive SMEs of homodyne and heterodyne detection rest on the limit of a strong local oscillator, in which individual photocounts are replaced by a continuous photocurrent~\cite{WisemanMilburn1993}. They therefore do not capture the discreteness of the counts, which can matter for weak local oscillators or short integration times.
	
	Collision models, also known as repeated-interaction models, provide an alternative route to open-system dynamics~\cite{Ciccarello2022,CampbellVacchini2021}. In such models, the system interacts sequentially with a stream of ancillas, each prepared in a known state and discarded after a short interaction. Tracing out the ancillas yields a discrete-time map on the system, and the Lindblad master equation emerges in the limit of short collision times~\cite{AttalPautrat2006,Ciccarello2022}. Because each ancilla can be measured after its collision, collision models also provide a transparent picture of quantum trajectories, in which the stochastic evolution follows from the Born rule applied to every ancilla~\cite{Brun2002,GoughSobolev2004,Pellegrini2008,Bouten2009}. Their discrete structure moreover makes them straightforward to simulate numerically.
	
	Several works have exploited this idea for the detection of propagating fields. Discrete models in which the field is represented by a chain of qubits reproduce the stochastic Schr\"odinger equations of photon counting and homodyne detection in the continuum limit~\cite{GoughSobolev2004,Pellegrini2008,Bouten2009}, and qubit probe models recover the standard SMEs of weak continuous measurement, including homodyne detection~\cite{Brun2002,Gross2018}. Collisional treatments with bosonic ancillas describe homodyne and heterodyne detection as projections of the ancillas onto quadrature or coherent states~\cite{Albarelli_2024}. In these models, the detection chain is replaced by an idealized measurement of the probe, although heterodyne detection has also been constructed by splitting a qubit probe on a beam splitter whose second port is in its ground state~\cite{Gross2018}. Formulations of input--output theory in terms of time bins or quantum pulses give closely related discrete descriptions, which focus on the state of the emitted field rather than on its subsequent processing and detection~\cite{Ciccarello2017,Fischer2018,KiilerichMolmer2019}. In parallel, the theory of cascaded quantum systems~\cite{Gardiner1993,Carmichael1993PRL} has been extended to general quantum networks in the SLH formalism, in which each component is specified by a scattering matrix $S$, a vector of coupling operators $L$, and a Hamiltonian $H$~\cite{GoughJames2009,Combes2017}. This formalism allows networks of quantum components, including beam splitters and phase shifters, to be assembled in a modular fashion, and complete parametrizations of diffusive unravelings have identified networks of beam splitters, phase shifters, and homodyne detectors that implement a given unraveling~\cite{ChiaWiseman2011}. These network descriptions, however, are formulated in continuous time and rest on quantum stochastic calculus. To our knowledge, no collision model has been developed in which the detection chain itself, including its local oscillators, vacuum ports, and photodetectors, is built from gates and click operators acting on bosonic ancillas, so that the Kraus operators of one detection scheme can be reused as a module in another.
	
	In this work, we develop such a modular linear-optics collision model for the readout of propagating fields. In each collision, the field emitted by the system during one time step enters a linear optical circuit that represents the complete detection apparatus and terminates in photodetectors. The auxiliary inputs of this circuit, such as the local oscillator and unused vacuum ports, are bosonic ancilla modes prepared in known states, and each optical element is represented by a gate unitary acting on the relevant modes~\cite{Leonhardt1997,Weedbrook2012}. The construction rests on a simple observation. Because the mean number of photons per collision is proportional to the collision time, at most one photon reaches each detector in a short collision, so that the Kraus operator of a click is simply the annihilation operator of the detected mode. Each gate of the chain transforms this operator linearly into a combination of field and ancilla operators. Tracing out the ancillas, which for vacuum and coherent states amounts to a projection, then yields effective Kraus operators that act on the system alone but encode the entire detection chain. What distinguishes the present approach is its modularity. Adding an element to the chain amounts to inserting one additional gate, and the effective Kraus operators of a simpler detection scheme serve directly as a module for a more complex one. Like earlier discrete treatments~\cite{Brun2002,Gross2018,Albarelli_2024}, the construction requires no stochastic calculus. In addition, it generates single measurement trajectories that retain individual photocounts, and in the continuous-time limit, it reduces to the standard master equation and, for a strong local oscillator, gives results consistent with standard SMEs.
	
	We highlight this modularity with two examples. First, we derive the effective Kraus operators of a homodyne module and use this module, together with one additional beam splitter, to construct the Kraus operators of heterodyne detection. As consistency checks, the homodyne Kraus operators coincide with the measurement operators of the standard description, in which the system output is photodetected after being mixed with a coherent local oscillator~\cite{WisemanMilburn1993,Carmichael1993,WisemanMilburn2009}, and the heterodyne module reproduces the reduction of the signal per quadrature by a factor of $\sqrt{2}$ relative to homodyne detection, which is the familiar 3-dB penalty~\cite{YamamotoHaus1986,CavesDrummond1994,Clerk2010}. We then apply the heterodyne module to dispersive qubit readout, treating the qubit and the readout mode jointly. This reveals a fundamental assignment error that depends on the initial qubit state and is present at any finite integration time even in an ideal readout chain, i.e., without qubit decoherence, photon loss, or classical noise. Because the model retains the discreteness of the photocounts, the heterodyne outcomes in the IQ plane also cross over from a Skellam distribution~\cite{Skellam1946} to a Gaussian distribution as the local-oscillator photon flux increases, as expected from earlier analyses of balanced homodyne detection with weak local oscillators~\cite{VogelGrabow1993,TycSanders2004}. Second, we add phase-preserving gain by replacing only the collision gate, while the heterodyne module is reused unchanged. The new three-mode gate combines a beam-splitter coupling of the readout mode to its output line with two-mode squeezing between the output line and an idler line, as in parametric couplers used for nonreciprocal amplification~\cite{MetelmannClerk2015,Sliwa2015,Lecocq2017}. The gain adds the minimum noise allowed for a phase-preserving amplifier~\cite{Caves1982,Clerk2010}, which can be read off directly from the resulting Kraus operators. Because the gain acts within the coupling, it also acts back on the readout mode, enhancing its decay rate and, for a coherent idler, driving it. This back-action must be taken into account to obtain the correct signal-to-noise ratio (SNR). The more common configuration, in which a circulator isolates the readout resonator from an amplifier placed in its output line, requires an amplifier of finite bandwidth, which we leave for future work.
	
	Because each element of a detection chain enters as an independent gate, the framework provides a transparent and extensible tool for modeling and optimizing realistic readout chains. Since it involves only gate unitaries, Kraus operators, and the Born rule, we expect it to also offer an accessible route to continuous measurements of propagating fields, complementing earlier discrete and pedagogical treatments~\cite{Brun2002,Gross2018,Albarelli_2024}.
	
	The paper is organized as follows. Section~\ref{sec:level2} reviews the collision model and describes how effective Kraus operators are constructed from linear-optics gates. Section~\ref{sec:level3} derives the homodyne Kraus operators, compares them with the standard SME description, and composes them into a heterodyne module. Section~\ref{sec:level4} introduces the dispersive readout model, applies the heterodyne module to it, and analyzes the finite-time assignment error and the Skellam-to-Gaussian crossover. Section~\ref{sec:level5} includes phase-preserving gain in the collision, derives the resulting Kraus operators, computes the SNR, and analyzes the back-action on the readout mode with the resulting master equation. Section~\ref{sec:level6} summarizes our results and gives an outlook, and Appendix~\ref{app:snr} derives the SNR of the gain--heterodyne readout.

	\begin{figure*}[t]
		\includegraphics[width=\textwidth]{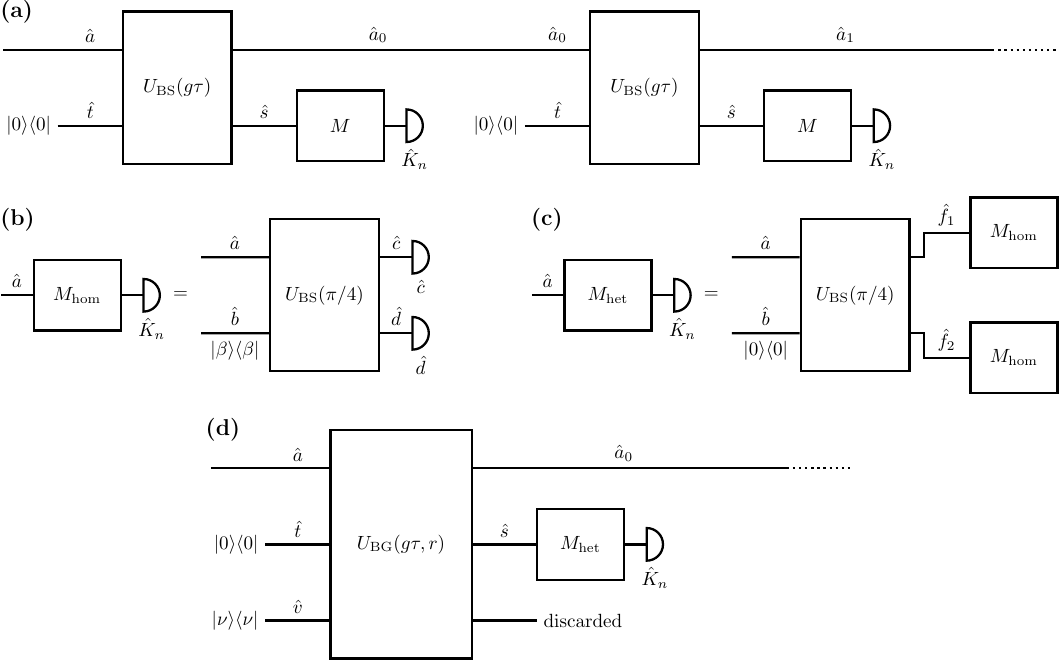}
		
		\caption{Collision model and measurement schemes $M$.
			(a)~Collision model. At each step, the system mode $\hat a$ interacts with a fresh ancilla $\hat t$ in the state $|0\rangle\langle0|$ through the beam-splitter unitary $U_{\mathrm{BS}}(g\tau)$, see Eqs.~\eqref{eq:U_BS} and \eqref{eq:U_BS_coll}. The system output ($\hat a_1$, $\hat a_2,\dots$) is passed on to the next collision, while the ancilla output $\hat s$ is measured by a device $M$ with Kraus operators $\hat K_k$, where $k$ labels the outcome. (b)~Homodyne measurement. The ancilla output $\hat s$ is mixed with the local-oscillator mode $\hat b$, prepared in the coherent state $|\beta'\rangle\langle\beta'|$, on a balanced beam splitter $U_{\mathrm{BS}}(\pi/4)$, and both outputs are detected, with click operators $\hat K_1=\hat c$ and $\hat K_2=\hat d$. (c)~Heterodyne measurement. The ancilla output $\hat s$ is mixed with the vacuum $|0\rangle\langle0|$ on a balanced beam splitter, and the two outputs $\hat f_1$ and $\hat f_2$ are measured by the homodyne measurements $M_{\mathrm{hom}}^{I}$ and $M_{\mathrm{hom}}^{Q}$ of the I and Q quadratures, respectively. (d)~Collision with gain. The system mode $\hat a$ interacts with a fresh readout bin $\hat t$ in the vacuum and a fresh idler bin $\hat v$ in the coherent state $|\nu\rangle$ through the three-mode unitary $U_{\mathrm{BG}}(g\tau,r)$ of Sec.~\ref{sec:gain_collision}. The output $\hat s$ of the readout bin is measured by a heterodyne measurement $M_{\mathrm{het}}$, and the idler bin is discarded.}
		\label{fig:all-gates}
	\end{figure*}
	
	
	\section{\label{sec:level2}Collision model and effective Kraus operators}
	
	In this section, we review the collision model and its connection to generalized measurements (Sec.~\ref{sec:coll_general}), discuss the scaling that yields the continuous-time limit (Sec.~\ref{sec:continuum}), and describe how a detection chain built from linear-optics gates is reduced to effective Kraus operators on the system (Secs.~\ref{sec:gates} and \ref{sec:coll_bs}).
	
	\subsection{\label{sec:coll_general}Repeated interactions and generalized measurements}
	
	In a collision model, also called a repeated-interaction model, a system of interest interacts with ancilla degrees of freedom that are initially prepared in a chosen state~\cite{Ciccarello2022,CampbellVacchini2021,Ciccarello2017,Cusumano2022}. The interaction lasts for a time $\tau$, after which the ancilla is either measured projectively or simply discarded. A fresh ancilla, prepared in the same state, then interacts with the system for another time $\tau$, or the same ancilla is reset to its initial state, and the process is repeated. The interaction Hamiltonian, and hence the corresponding unitary, is the same at every step. Collision models date back to the work of Rau on relaxation phenomena~\cite{Rau1963} and were later used to describe thermalization~\cite{Scarani2002}. They now provide a versatile framework for open-system dynamics, quantum optics, and quantum trajectories, as reviewed in Refs.~\cite{Ciccarello2022,CampbellVacchini2021,Ciccarello2017,Cusumano2022}.
	
	Let $\hat U=\exp[-i(\hat H_S+\hat H_{SA}+\hat{H}_A)\tau]$ be the unitary of one collision, where $\hat H_S$ is the system Hamiltonian, $\hat H_{SA}$ is the system--ancilla interaction, and $\hat H_A$ is the Hamiltonian of the ancilla (we set $\hbar=1$). For an ancilla prepared in a pure state $\ket{\psi}$, as is the case for the vacuum and coherent-state ancillas considered in this work, tracing out the ancilla yields a completely positive, trace-preserving map in Kraus form~\cite{Kraus1983,Nielsen_Chuang_2010,Ciccarello2022},
	\begin{equation}
		\begin{aligned}
			\mathrm{Tr}_A\!\left[\hat U\big(\hat\rho_S\otimes|\psi\rangle\langle\psi|\big)\hat U^\dagger\right]
			&=\sum_k \hat K_k\,\hat\rho_S\,\hat K_k^\dagger,\\
			\hat K_k&=\langle k|\hat U|\psi\rangle,
		\end{aligned}
		\label{eq:coll_map}
	\end{equation}
	where $\{\ket{k}\}$ is an orthonormal basis of the ancilla. The matrix element that defines $\hat K_k$ is taken over the ancilla only, so that $\hat K_k$ acts on the system, and unitarity of $\hat U$ guarantees the completeness relation $\sum_k\hat K_k^\dagger\hat K_k=\langle\psi|\hat U^\dagger\hat U|\psi\rangle=\hat{\mathbb{I}}$.
	
	If the ancilla is measured projectively in the basis $\{\ket{k}\}$ after the collision, Eq.~\eqref{eq:coll_map} acquires an operational meaning. By the Born rule, outcome $k$ occurs with probability $p_k=\mathrm{Tr}(\hat K_k\hat\rho_S\hat K_k^\dagger)=\mathrm{Tr}(\hat E_k\hat\rho_S)$, with $\hat E_k=\hat K_k^\dagger\hat K_k$, and the system is updated to the conditional state $\hat K_k\hat\rho_S\hat K_k^\dagger/p_k$. Each collision therefore implements a generalized measurement on the system, described by the positive-operator-valued measure (POVM) $\{\hat E_k\}$, with the ancilla acting as the meter~\cite{Nielsen_Chuang_2010,WisemanMilburn2009,JacobsSteck2006}. Averaging over the outcomes recovers the unconditional map of Eq.~\eqref{eq:coll_map}. Over many collisions, the outcomes $(k_1,\dots,k_n)$ form a measurement record, and the conditional state
	\begin{equation}
		\hat\rho_S^{(n)}\propto\hat K_{k_n}\cdots\hat K_{k_1}\,\hat\rho_S^{(0)}\,\hat K_{k_1}^\dagger\cdots\hat K_{k_n}^\dagger
		\label{eq:discrete_traj}
	\end{equation}
	traces out a discrete quantum trajectory~\cite{Brun2002,Bouten2009,Gross2018,Ciccarello2022}. Measuring the ancillas in a different basis transforms the Kraus operators by a unitary mixing that leaves the unconditional map of Eq.~\eqref{eq:coll_map} invariant~\cite{Nielsen_Chuang_2010}, so different measurements correspond to different unravelings of the same unconditional dynamics~\cite{WisemanMilburn2009,Brun2002}. In the present work, this choice of measurement is made by the detection chain that processes the ancilla output before photodetection. Because the ancillas are initially uncorrelated and each interacts with the system only once, the resulting dynamics is Markovian, whereas initially correlated ancillas or ancilla--ancilla collisions give rise to non-Markovian dynamics~\cite{Ciccarello2022,CampbellVacchini2021,Cilluffo_2019}.
	
	\subsection{\label{sec:continuum}Continuous-time limit and the $\sqrt{\tau}$ scaling}
	
	A continuous measurement emerges from a sequence of weak collisions in the limit $\tau\to0$~\cite{AttalPautrat2006,Brun2002,GoughSobolev2004,Pellegrini2008,Bouten2009,Gross2018,Albarelli_2024}. For direct photodetection of the ancillas, $\tau$ then plays the role of the infinitesimal time step $dt$ of the SME, whereas for the diffusive unravelings of Sec.~\ref{sec:level3}, $dt$ corresponds to a coarse-grained interval that contains many collisions. The limit $\tau\to0$ requires care. To obtain a finite rate of dissipation, and hence of information gain, the system--ancilla coupling cannot be held fixed but must diverge as $\tau^{-1/2}$~\cite{AttalPautrat2006,Ciccarello2017,Ciccarello2022,Cusumano2022,Gross2018}. We illustrate this for the situation of interest here, in which a bosonic system mode $\hat a$ with Hamiltonian $\hat H_S$ exchanges excitations with a vacuum ancilla mode $\hat t$ [Fig.~\ref{fig:all-gates}(a)]. For two bosonic modes $\hat x$ and $\hat y$, we denote by
	\begin{equation}
		U_{\mathrm{BS}}(\theta)=\exp\!\big[-i\theta\big(\hat x^\dagger\hat y+\hat x\hat y^\dagger\big)\big]
		\label{eq:U_BS}
	\end{equation}
	the beam-splitter unitary with mixing angle $\theta$, which maps $\hat x\to\hat x\cos\theta-i\hat y\sin\theta$ and $\hat y\to\hat y\cos\theta-i\hat x\sin\theta$ in the Heisenberg picture. The collision is the beam splitter between $\hat a$ and $\hat t$ generated by an exchange interaction of strength $g$ that acts for the time $\tau$,
	\begin{equation}
		U_{\mathrm{BS}}(g\tau)=e^{-i\tau g\hat H_{\mathrm{BS}}},\qquad
		\hat H_{\mathrm{BS}}=\hat a^\dagger\hat t+\hat a\hat t^\dagger ,
		\label{eq:U_BS_coll}
	\end{equation}
	i.e., Eq.~\eqref{eq:U_BS} with $(\hat x,\hat y)=(\hat a,\hat t)$ and mixing angle $\theta=g\tau$; in the notation of Sec.~\ref{sec:coll_general}, $\hat H_{SA}=g\hat H_{\mathrm{BS}}$. We set $\hat H_A=0$ and include $\hat H_S$ by interleaving its free evolution with the collisions, which agrees with $\hat U$ of Sec.~\ref{sec:coll_general} to first order in $\tau$. For a measurement of the ancilla photon number, and to lowest nontrivial order in $\tau$ and $g\tau$, the Kraus operators of Eq.~\eqref{eq:coll_map} for zero and one detected photon are~\cite{Ciccarello2022,Gross2018}
	\begin{equation}
		\begin{aligned}
			\hat K_0&\simeq\hat{\mathbb{I}}-i\hat H_S\tau-\tfrac12(g\tau)^2\,\hat a^\dagger\hat a,\\
			\hat K_1&\simeq g\tau\,\hat a,
		\end{aligned}
		\label{eq:coll_kraus}
	\end{equation}
	up to an irrelevant phase of $\hat K_1$, while outcomes with two or more photons contribute only at higher order. One collision therefore maps the system state to
	\begin{equation}
		\begin{aligned}
			\hat\rho_S\to\hat\rho_S&-i\tau[\hat H_S,\hat\rho_S]\\
			&+(g\tau)^2\Big(\hat a\hat\rho_S\hat a^\dagger-\tfrac12\big\{\hat a^\dagger\hat a,\hat\rho_S\big\}\Big).
		\end{aligned}
		\label{eq:coll_step}
	\end{equation}
	
	The Hamiltonian term is of order $\tau$, whereas the dissipative term is of second order in the mixing angle. Since collisions occur at a rate $1/\tau$, a finite decay rate $\kappa$ in the limit $\tau\to0$ requires $(g\tau)^2=\kappa\tau$, that is,
	\begin{equation}
		g=\sqrt{\kappa/\tau}\;\propto\;\tau^{-1/2}.
		\label{eq:sqrt_scaling}
	\end{equation}
	This is the usual scaling of collision models, required for the continuous-time limit~\cite{AttalPautrat2006,Ciccarello2017,Ciccarello2022}. With it, the mixing angle of the collision is $g\tau=\sqrt{\kappa\tau}$, so that the decay rate $\kappa$ enters the model through the collision unitary, both terms in Eq.~\eqref{eq:coll_step} are of order $\tau$, and the limit $\tau\to0$ yields the Lindblad master equation~\cite{Lindblad1976,GKS1976,AttalPautrat2006,Ciccarello2022}
	\begin{equation}
		\begin{aligned}
			\dot{\hat\rho}_S={}&-i[\hat H_S,\hat\rho_S]\\
			&+\kappa\Big(\hat a\hat\rho_S\hat a^\dagger-\tfrac12\big\{\hat a^\dagger\hat a,\hat\rho_S\big\}\Big).
		\end{aligned}
		\label{eq:coll_lindblad}
	\end{equation}
	A coupling that grows more slowly than $\tau^{-1/2}$ produces no dissipation in this limit, whereas a faster growth leads to a divergent rate. More generally, when the ancilla output is registered by several detectors whose clicks are described by Kraus operators $\hat K_j$ of order $\sqrt{\tau}$, the no-click operator $\hat K_0=\hat{\mathbb{I}}-i\hat H_S\tau-\tfrac12\sum_j\hat K_j^\dagger\hat K_j$ ensures completeness to first order in $\tau$~\cite{WisemanMilburn2009,JacobsSteck2006}.
	
	With Eq.~\eqref{eq:sqrt_scaling}, $\hat K_1=\sqrt{\kappa\tau}\,\hat a$. The probability of a detection event in one collision, $p_1=\kappa\tau\langle\hat a^\dagger\hat a\rangle$, is thus of order $\tau$, whereas its amplitude is of order $\sqrt{\tau}$. Terms of second order in the collision amplitude therefore contribute at first order in $\tau$. This is the collisional counterpart of the It\^o rule $dW^2=dt$ for a Wiener increment and of the It\^o rules of quantum stochastic calculus~\cite{HudsonParthasarathy1984,GardinerCollett1985,GardinerZoller2004,JacobsSteck2006}. The physical origin of the scaling becomes transparent when the input field $\hat b_{\mathrm{in}}(t)$, with $[\hat b_{\mathrm{in}}(t),\hat b_{\mathrm{in}}^\dagger(t')]=\delta(t-t')$, is discretized into time bins of duration $\tau$~\cite{Ciccarello2017,Ciccarello2022,Fischer2018}. The bin operators
	\begin{equation}
		\hat t_n=\frac{1}{\sqrt{\tau}}\int_{t_n}^{t_n+\tau}\hat b_{\mathrm{in}}(t)\,dt
		\label{eq:time_bins}
	\end{equation}
	satisfy $[\hat t_n,\hat t_m^\dagger]=\delta_{nm}$, so each bin acts as an independent bosonic ancilla. The noise increment $\Delta\hat B_n=\sqrt{\tau}\,\hat t_n$ satisfies $\langle\Delta\hat B_n\Delta\hat B_n^\dagger\rangle=\tau$ in the vacuum, which is the discrete analog of the It\^o rule $d\hat B\,d\hat B^\dagger=dt$. Integrated over one bin, the input--output coupling of strength $\sqrt{\kappa}$ becomes a beam splitter between $\hat a$ and $\hat t_n$ with mixing angle $g\tau=\sqrt{\kappa\tau}$, which is Eq.~\eqref{eq:sqrt_scaling}. The $\sqrt{\tau}$ scaling enters the system dynamics through the collision unitary. The same time-bin scaling fixes the amplitude per bin of a coherent ancilla that reaches the detectors, such as the local oscillator (Sec.~\ref{sec:homodyne}). All subsequent steps involve only gates, Kraus operators, and the Born rule.
	
	Direct photodetection of the ancillas generates the quantum-jump unraveling of Eq.~\eqref{eq:coll_lindblad}~\cite{Dalibard1992,Gardiner1992,Carmichael1993}. If the ancilla output is instead mixed with a strong local oscillator before photodetection, many detection events accumulate within any coarse-grained time interval, and the trajectories become diffusive, as in homodyne and heterodyne detection~\cite{WisemanMilburn1993,WisemanDiosi2001,ChiaWiseman2011,Genoni_2014,Brun2002,Gross2018}. In the present work, the detection chain acts on the system only through the collisional beam splitter $U_{\mathrm{BS}}(g\tau)$, whose small mixing angle $g\tau=\sqrt{\kappa\tau}$ makes each collision a weak measurement of the system~\cite{Brun2002,Ciccarello2022,AttalPautrat2006}. The ancilla output $\hat s$ enters a measurement device $M$ that represents the complete detection chain [Figs.~\ref{fig:all-gates}(b) and \ref{fig:all-gates}(c)], and in Sec.~\ref{sec:level5}, the collision itself is extended to include gain [Fig.~\ref{fig:all-gates}(d)]. The remainder of this section describes how the Kraus operators of $M$ are constructed and transferred to the system.
	
	\subsection{\label{sec:gates}Detection chains as linear-optics gates}
	
\begin{figure}[htbp]
		\centering
		\begin{tikzpicture}[
			gate/.style={draw, thick, minimum width=2.2cm, minimum height=3cm, fill=white},
			line/.style={thick},
			every node/.style={font=\normalsize}
			]
			
			\draw[line] (0,2.0) -- (1.4,2.0);
			\node[above] at (0.7,2.0) {$\hat a$};
			\node[left] at (0,2.0) {$\hat \rho_S$};
			
			\draw[line] (0,0.4) -- (1.4,0.4);
			\node[below] at (0.7,0.4) {$\hat b$};
			\node[left] at (0,0.4) {$|\psi\rangle\langle\psi|$};
			
			\node[gate] (G1) at (2.4,1.2) {$U$};
			
			\draw[line] (3.5,2.0) -- (4.9,2.0);
			\node[above] at (4.2,2.0) {$\hat c$};
			
			\draw[line] (3.5,0.4) -- (4.9,0.4);
			\node[below] at (4.2,0.4) {$\hat d$};
			
			\node[right] at (5.0,2.0) {$\hat K_1=\;\hat c$};
			\node[right] at (5.0,0.4) {$\hat K_2=\;\hat d$};
			
		\end{tikzpicture}
		\caption{General two-mode gate. The signal $\hat{\rho}_S$ (mode $\hat a$) and an ancilla in the known state $\ket{\psi}$ (mode $\hat b$) enter the unitary $U$, and the outputs $\hat{c}$ and $\hat{d}$ are detected with Kraus operators $\hat{K}_1=\hat{c}$ and $\hat{K}_2=\hat{d}$.}
		\label{fig:general-gate}
	\end{figure}
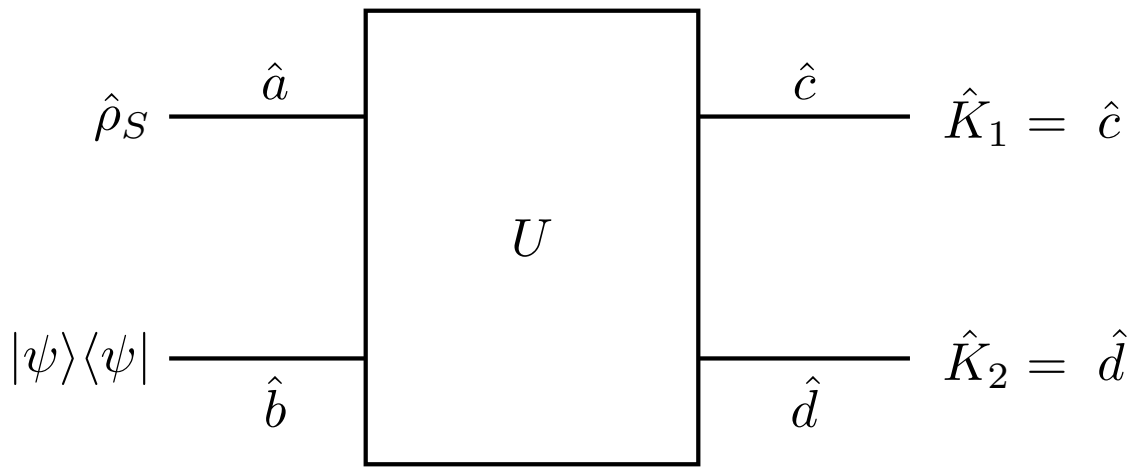

	We consider measurement schemes assembled from passive linear-optics elements, such as beam splitters and phase shifters, which act as gates on bosonic modes~\cite{Leonhardt1997,Weedbrook2012}. Since any passive linear-optics network can be decomposed into beam splitters and phase shifters~\cite{PhysRevLett.73.58}, it suffices to consider gates that act on at most two modes. A two-mode gate maps the annihilation operators $(\hat{a},\hat{b})$ of its input modes linearly onto the annihilation operators $(\hat{c},\hat{d})$ of its output modes,
	\begin{equation}
		\begin{bmatrix}
			\hat{c} \\
			\hat{d}
		\end{bmatrix}
		=
		\begin{bmatrix}
			U_{00} & U_{01} \\
			U_{10} & U_{11}
		\end{bmatrix}
		\begin{bmatrix}
			\hat{a} \\
			\hat{b}
		\end{bmatrix},
		\label{eq:gen_uni_mat}
	\end{equation}
	where $(U_{ij})$ is a unitary matrix. In the detection chains considered here, one input of each gate carries the signal, and the other is an ancilla in a known state, such as the vacuum or a coherent local oscillator (Fig.~\ref{fig:general-gate}). A cascade of such gates terminates in photodetectors, each of which registers a click or no click in every collision. Provided that at most one photon reaches a detector per collision, which we verify in Sec.~\ref{sec:homodyne}, a click is described by a Kraus operator equal to the annihilation operator of the detected mode.
	
	Starting from the click operators of the detectors, we work backward through the chain and obtain, gate by gate, effective Kraus operators on the input modes, until we arrive at Kraus operators that act only on the Hilbert space of the measured mode~\cite{Leonhardt1997,Weedbrook2012}. For a single gate, this amounts to expressing the click operators of the output modes $\hat{c}$ and $\hat{d}$ as functions of the input operators $\hat{a}$ and $\hat{b}$ by means of Eq.~\eqref{eq:gen_uni_mat}, and then projecting the ancilla mode $\hat{b}$ onto its known input state $\ket{\psi}$ (Fig.~\ref{fig:general-gate})~\cite{Nielsen_Chuang_2010},
	\begin{equation}
		\hat{K}_{\hat{a}} = (\hat{\mathbb{I}}\otimes\langle\psi|)\,\hat{K}_{1/2}(\hat{a}, \hat{b})\,(\hat{\mathbb{I}}\otimes|\psi\rangle).
		\label{eq:gen_uni_eq}
	\end{equation}
	Here, $\hat{K}_{1/2}(\hat{a},\hat{b})$ is the click operator $\hat K_1=\hat{c}$ or $\hat K_2=\hat{d}$ of Fig.~\ref{fig:general-gate}, expressed through Eq.~\eqref{eq:gen_uni_mat} as a linear combination of $\hat{a}$ and $\hat{b}$. The projection in Eq.~\eqref{eq:gen_uni_eq} is exact whenever the click operator contains $\hat{b}$ but not $\hat{b}^\dagger$ and $\ket{\psi}$ is an eigenstate of $\hat b$, as is the case for a coherent state or the vacuum. A click then leaves the ancilla in its initial state, so the ancilla does not become entangled with the measured mode, and $\hat b$ can be replaced by its eigenvalue. Otherwise, the final state of the ancilla must be resolved in an orthonormal basis, following Eq.~\eqref{eq:coll_map}, as we do for the idler of the collision with gain in Sec.~\ref{sec:level5}.
	
	The resulting Kraus operators act on the measured mode alone, so the conditional dynamics can be simulated on a single Hilbert space. Modeling every mode of the chain explicitly would instead require a joint Hilbert space whose dimension grows exponentially with the number of modes.
	
	\subsection{\label{sec:coll_bs}Collisional beam splitter}
	
	For the collisional beam splitter of Fig.~\ref{fig:all-gates}(a), Eq.~\eqref{eq:U_BS_coll} gives the output mode $\hat{a}_1$, which is passed on to the next collision, and the output mode $\hat{s}$, which enters the detection chain,
	\begin{equation}
		\begin{aligned}
			\hat{a}_1 &= \hat{a}\cos{g\tau}-i\hat{t}\sin{g\tau},\\
			\hat{s} &= -i\hat{a}\sin{g\tau}+\hat{t}\cos{g\tau}.
		\end{aligned}
		\label{eq:operator_scaling}
	\end{equation}
	Projecting the vacuum ancilla mode $\hat{t}$ onto $\ket{0}$ as in Eq.~\eqref{eq:gen_uni_eq} and expanding to lowest order in $g\tau$ gives
	\begin{equation}
		\begin{aligned}
			\hat{a}_1 &= \hat{a},\\
			\hat{s} &= -ig\tau\,\hat{a},
		\end{aligned}
		\label{eq:operator_scaling_2}
	\end{equation}
	where $g\tau=\sqrt{\kappa\tau}$ by Eq.~\eqref{eq:sqrt_scaling}. The collisional beam splitter and the balanced beam splitters of the detection chain are thus instances of the same gate, Eq.~\eqref{eq:U_BS}, with mixing angles $\theta=g\tau\ll1$ and $\theta=\pi/4$, respectively. The decay rate $\kappa$ enters the model only through the small mixing angle of the collision, whereas the photodetectors at the end of the detection chain are described by click operators equal to the annihilation operators of the detected modes, without any prefactor (see later in Sec.~\ref{sec:homodyne}). The phase $-i$ of $\hat{s}$ relative to $\hat{a}$ can be absorbed into the local-oscillator phase, where it only rotates the measured quadrature, so we omit it in what follows and use $\hat{s}=\sqrt{\kappa\tau}\,\hat{a}$.
	
	Because $\kappa$ enters only through $\hat s$, the role of the collision can be stated for any detection chain. Since the chain consists of linear gates and its ancillas are projected onto eigenstates of their annihilation operators, the effective Kraus operator for a click of each detector has the affine form $\hat{K}_{\hat s}=A(B\hat{s}+C\hat{\mathbb{I}})$ on the Hilbert space of $\hat{s}$, where the complex constants $A$ and $B$ are set by the gates and $C$ is proportional to the ancilla amplitudes, such as that of the local oscillator. Substituting $\hat{s}=\sqrt{\kappa\tau}\,\hat{a}$ gives the corresponding Kraus operator on the Hilbert space of $\hat{a}$, $\hat{K}_{\hat a}=A\sqrt{\kappa\tau}\,(B\hat{a}+\tfrac{C}{\sqrt{\kappa\tau}}\hat{\mathbb{I}})$. The collision thus multiplies $A$ by $\sqrt{\kappa\tau}$, which makes each click probability of order $\tau$, and divides $C$ by $\sqrt{\kappa\tau}$. For $C/\sqrt{\kappa\tau}$ to remain finite and nonzero as $\tau\to0$, $C$ must therefore scale as $\sqrt{\kappa\tau}$, as we show next explicitly for the local oscillator.
	
	\section{\label{sec:level3}Homodyne and heterodyne detection}
	
	\subsection{\label{sec:sme}Standard SME description}
	
	To set the stage, we first summarize the standard SME description of homodyne and heterodyne detection. The unconditional dynamics of the system is generated by the Lindblad master equation [cf.\ Eq.~\eqref{eq:coll_lindblad}]
	\begin{equation}
		\begin{aligned}
			\dot{\hat\rho}&=\mathcal{L}\hat\rho=-i[\hat H,\hat\rho]
			+\sum_{k}\mathcal{D}[\hat L_k]\hat\rho,
			\\
			\mathcal{D}[\hat L]\hat\rho&=\hat L\hat\rho\hat L^{\dagger}
			-\tfrac12\{\hat L^{\dagger}\hat L,\hat\rho\},
		\end{aligned}
		\label{eq:lindblad}
	\end{equation}
	where $\hat L_k$ is the jump operator of output channel $k$ (for the system mode $\hat a$, $\hat L=\sqrt{\kappa}\,\hat a$). In photon counting, the term $\hat L_k\hat\rho\hat L_k^{\dagger}$ describes a click of an ideal detector on channel $k$, which occurs with probability $dt\,\langle\hat L_k^{\dagger}\hat L_k\rangle_c$ in the conditional state. Averaging over the click records recovers Eq.~\eqref{eq:lindblad}, so photon counting is one unraveling of the master equation. Photon counting is, however, insensitive to the phase of the emitted field and therefore gives no access to its quadratures.
	
	Homodyne and heterodyne detection correspond to different unravelings of the dissipator in Eq.~\eqref{eq:lindblad}~\cite{WisemanMilburn2009,WisemanMilburn1993,PRXQuantum.5.020201}. The master equation is invariant under the shift $\hat{L}_k\to\hat{L}_k+\beta_k\hat{\mathbb{I}}$, together with a compensating change of $\hat{H}$, so the clicks can equally be associated with the shifted jump operators $\hat{L}_k+\beta_k\hat{\mathbb{I}}$. Physically, the shift describes photodetection of the system output after it has been mixed with a local oscillator (LO) of amplitude $\beta_k$. In balanced homodyne detection, the system output and the LO are mixed on a balanced beam splitter, and the two detectors are associated with the jump operators $(\hat L_k\pm\beta_k\hat{\mathbb{I}})/\sqrt2$~\cite{WisemanMilburn1993,WisemanMilburn2009}. For large $|\beta_k|$, the click rate is dominated by the LO, and the jump process, coarse-grained over many clicks, is approximated by a Gaussian diffusive process. Its mean carries the system quadrature $\hat x_k=\hat L_k e^{-i\phi_k}+\hat L_k^\dagger e^{i\phi_k}$, with $\phi_k=\arg\beta_k$~\cite{PRXQuantum.5.020201}. The resulting conditional state obeys
	\begin{equation}
		\begin{aligned}
			d\hat\rho_c &= dt\,\mathcal{L}\hat\rho_c
			+ \sum_k dW_k\left[\mathcal{H}_k\hat\rho_c - \langle \hat x_k\rangle_c\hat\rho_c\right],
			\\
			\mathcal{H}_k\hat\rho &= \hat L_k e^{-i\phi_k}\hat\rho + \hat\rho \hat L_k^\dagger e^{i\phi_k},
		\end{aligned}
	\end{equation}
	with independent Wiener increments $dW_k$, and the measured current is $J_k = w_k(\langle \hat x_k\rangle_c + dW_k/dt)$, where $w_k$ is a scale factor that normalizes the signal.
	
	For homodyne detection, $\hat{L}_k=\sqrt{\kappa}\,\hat{a}$ and $w_k=1/\sqrt{2\kappa}$, so that the current has mean $\langle\hat{x}_\phi\rangle$, with $\hat{x}_\phi=(\hat{a}e^{-i\phi}+\hat{a}^\dagger e^{i\phi})/\sqrt{2}$. Heterodyne detection splits the output on a balanced beam splitter and performs homodyne detection in each arm, with $\hat{L}_1=\hat{L}_2=\sqrt{\kappa/2}\,\hat{a}$ and reference phases $0$ and $\pi/2$. Here, $w_1=w_2=\sqrt{1/\kappa}$, which is larger by a factor of $\sqrt{2}$ than in the homodyne case. This larger normalization is the price of measuring two conjugate quadratures simultaneously, since vacuum fluctuations enter through the unused port of the beam splitter~\cite{WisemanMilburn2009,PRXQuantum.5.020201}. Equivalently, at fixed noise, the signal per quadrature is reduced by a factor of $\sqrt{2}$, which is the familiar 3-dB penalty of heterodyne detection~\cite{YamamotoHaus1986,CavesDrummond1994,Clerk2010}. 
	
	This description has two limitations, which motivate the modular construction presented below. First, the diffusive SME is obtained in the joint limit of continuous time, $dt\to0$, and a strong local oscillator, $|\beta_k|\to\infty$, in which many clicks occur within each $dt$. It therefore cannot describe a weak local oscillator, for which the discreteness of the photocounts matters (Sec.~\ref{sec:skellam}). Second, each element of the detection chain has to be translated by hand into modified jump operators or effective parameters of the SME, such as an efficiency or added noise, unless the chain is treated within a continuous-time network formalism~\cite{GoughJames2009,Combes2017}.
	
	\subsection{\label{sec:our_approach}Our modular collision model description}
	
	In our approach, a detection scheme is assembled from the linear-optics gates of Sec.~\ref{sec:gates}, and its effective Kraus operators are obtained by propagating the click operators of the photodetectors back to the measured mode. We first derive the Kraus operators of homodyne detection, which allows a direct comparison with the standard description of Sec.~\ref{sec:sme}, and then use them as a module for heterodyne detection.
	
	\subsubsection{\label{sec:homodyne}Homodyne detection}
	
	In homodyne detection, shown in Fig.~\ref{fig:all-gates}(b), the mode to be measured and a local oscillator in mode $\hat b$ are mixed on a balanced beam splitter, and both output modes are registered by photodetectors~\cite{WisemanMilburn1993,Leonhardt1997,BanaszekWodkiewicz1997}. In the collision model, the mode entering the detection chain is not the system mode $\hat a$ itself but the ancilla output $\hat s$ of Eq.~\eqref{eq:operator_scaling}. Because the mixing angle of the collision is $g\tau=\sqrt{\kappa\tau}$ (Sec.~\ref{sec:continuum}), this mode is $\hat s=\sqrt{\kappa\tau}\,\hat a$ [Eq.~\eqref{eq:operator_scaling_2}], up to the phase omitted in Sec.~\ref{sec:coll_bs}. In each time bin of duration $\tau$, the local oscillator is a coherent state $|\beta'\rangle$ of mode $\hat b$, whose scaling with $\tau$ we determine below.
	
	
	
	
	
	
	
	
	
	The balanced beam splitter $U_{\mathrm{BS}}(\pi/4)$ of Eq.~\eqref{eq:U_BS}, acting on $\hat s$ and $\hat b$, transforms the input modes as
	\begin{equation}
		\begin{bmatrix}\hat c\\ \hat d\end{bmatrix}
		=
		\frac{1}{\sqrt{2}}
		\begin{bmatrix}1 & -i\\ 
			-i & 1
		\end{bmatrix}
		\begin{bmatrix}\hat s\\ \hat b\end{bmatrix},
		\qquad \theta=\pi/4,
		\label{eq:eq_balanced_bm}
	\end{equation}
	and the two output modes are registered by photodetectors with click operators $\hat K_1=\hat c=(\hat s-i\hat b)/\sqrt2$ and $\hat K_2=\hat d=(\hat b-i\hat s)/\sqrt2$. The click operator of each detector is the annihilation operator of the detected mode, without any prefactor. This form is exact for an ideal detector that absorbs the detected photon, provided that at most one photon reaches it: on the subspace with at most one photon, the annihilation operator represents the absorption $\ket{1}\to\ket{0}$ exactly. This condition holds in each collision, as we verify below, and the decay rate $\kappa$ enters only through $\hat s$. Since the local oscillator is in a known eigenstate of $\hat b$, it is projected out as in Eq.~\eqref{eq:gen_uni_eq}, $\hat K_{R_j}=(\hat{\mathbb I}\otimes\langle\beta'|)\,\hat K_j\,(\hat{\mathbb I}\otimes|\beta'\rangle)$, where $R\in\{I,Q\}$ labels the measured quadrature, using $\langle\beta'|\hat b|\beta'\rangle=\beta'$ and $\langle\beta'|\beta'\rangle=1$. After the substitution $\hat s=\sqrt{\kappa\tau}\,\hat a$, each effective Kraus operator combines the system operator $\sqrt{\kappa\tau}\,\hat a$, whose scaling is fixed by the collision, with the local-oscillator amplitude $\beta'$.
	
	The scaling of $\beta'$ with $\tau$ is not arbitrary but follows from the requirement of a continuous limit. The total click probability of the two detectors per time step is $\kappa\tau\langle\hat a^\dagger\hat a\rangle+|\beta'|^2$, whereas the quadrature signal arises from the interference between the system output and the local oscillator and is of order $\sqrt{\kappa\tau}\,|\beta'|$. A continuous limit with finite, nonzero rates requires both quantities to scale linearly with $\tau$. If $\beta'$ vanished faster than $\sqrt{\tau}$, the interference term would vanish in this limit, and the scheme would reduce to direct photodetection of the system output, without access to a quadrature. If $\beta'$ vanished more slowly than $\sqrt{\tau}$, or not at all, the click rate $|\beta'|^2/\tau$ would diverge as $\tau\to0$, and no continuous limit would exist. Hence, $\beta'\propto\sqrt{\tau}$. Writing the $\tau$-independent prefactor as $\sqrt{\kappa}\,\beta$, with a dimensionless amplitude $\beta$, we obtain
	\begin{equation}
		\beta'=\sqrt{\kappa\tau}\,\beta ,
		\label{eq:beta_scaling}
	\end{equation}
	i.e., $|\beta'|^2=\kappa\tau|\beta|^2$ photons per bin, which corresponds to the local-oscillator photon flux $\kappa|\beta|^2$. This is precisely the time-bin scaling of Sec.~\ref{sec:continuum}, but here it follows from the Kraus operators alone, without any stochastic calculus. With this scaling, the system operator and the local-oscillator amplitude carry the same factor $\sqrt{\kappa\tau}$, which factors out, and the homodyne Kraus operators for the clicks of the two detectors are
	\begin{equation}
		\begin{aligned}
			\hat{K}_{R_1} &=\sqrt{\tfrac{\kappa\tau}{2}}(\hat{a}-i\beta\hat{\mathbb{I}}),\\
			\hat{K}_{R_2} &=\sqrt{\tfrac{\kappa\tau}{2}}(\beta\hat{\mathbb{I}} - i\hat{a}),
		\end{aligned}
		\label{eq:homo_kraus}
	\end{equation}
	where the quadrature $R$ is selected by the local-oscillator phase $\phi$, defined by $\beta=|\beta|e^{i\phi}$. These Kraus operators act only on the Hilbert space of the mode that carries the signal, yet they represent the entire measurement apparatus, namely, the beam splitter, the local oscillator, and both detectors. Up to a global phase, they coincide with the measurement operators $\sqrt{\tau}\,(\sqrt{\kappa}\,\hat a\mp i\sqrt{\kappa}\,\beta)/\sqrt2$ of the two detectors in the standard description of balanced homodyne detection (Sec.~\ref{sec:sme})~\cite{WisemanMilburn1993,WisemanMilburn2009}. Here, they follow from the gates of the detection chain rather than from a shift of the jump operator, which provides a first consistency check of our approach. All subsequent expressions follow from Eq.~\eqref{eq:homo_kraus}.
	
	Each detector is described by a pair of Kraus operators, one for a click and one for no click. The no-click Kraus operator associated with $\hat K_{R_j}$ follows formally as $\hat K_{R_j}^{(0)}=\hat{\mathbb{I}}-\tfrac12\hat K_{R_j}^\dagger\hat K_{R_j}$, which satisfies $\hat K_{R_j}^{(0)\dagger}\hat K_{R_j}^{(0)}+\hat K_{R_j}^\dagger\hat K_{R_j}=\hat{\mathbb{I}}$ to first order in $\tau$. To this order, the product of the no-click operators of the two detectors and the free evolution $e^{-i\hat H_S\tau}$ equals the joint no-click operator $\hat K_0$ of Sec.~\ref{sec:continuum}. Averaging over the click and no-click outcomes of both detectors, the state after one time step is
	\begin{equation}
		\begin{aligned}
			\hat\rho(t+\tau)&=\hat\rho(t)+\sum_{j=1,2}\mathcal D[\hat K_{R_j}]\hat\rho(t)+\mathcal O(\tau^2)\\
			&=\hat\rho(t)+\kappa\tau\,\mathcal D[\hat a]\hat\rho(t)+\mathcal O(\tau^2),
		\end{aligned}
		\label{eq:homodyne_step}
	\end{equation}
	because $\mathcal D[\hat K_{R_{1,2}}]\hat\rho=\tfrac{\kappa\tau}{2}\mathcal D[\hat a]\hat\rho\pm\tfrac{i\kappa\tau}{4}[\beta^*\hat a+\beta\hat a^\dagger,\hat\rho]$, so that the local-oscillator terms of the two detectors cancel. This cancellation is expected, since a measurement performed on the ancilla output after the collision cannot change the unconditional dynamics of the system (Sec.~\ref{sec:coll_general}). In the limit $\tau\to0$, Eq.~\eqref{eq:homodyne_step} yields $\dot{\hat\rho}=\kappa\,\mathcal D[\hat a]\hat\rho$, which is the dissipative part of the Lindblad equation~\eqref{eq:lindblad} with $\hat L=\sqrt{\kappa}\,\hat a$ and the decay rate set by the collision. The Hamiltonian part follows by interleaving the unitary evolution with the measurement steps. Since every Kraus operator for a click has a no-click partner, we discuss only the click Kraus operators henceforth.
	
	The Kraus operators are derived to first order in $\tau$, consistent with the collision model of Sec.~\ref{sec:level2}. With Eq.~\eqref{eq:beta_scaling}, the total click probability of the two detectors per time step is $\kappa\tau(\langle\hat a^\dagger\hat a\rangle+|\beta|^2)$. The model therefore reproduces the correct measurement statistics only if
	\begin{equation}
		\kappa\tau\big(\langle\hat a^\dagger\hat a\rangle+|\beta|^2\big) \ll 1 ,
		\label{eq:dt_condition}
	\end{equation}
	which for a strong local oscillator, $|\beta|^2\gg\langle\hat a^\dagger\hat a\rangle$, reduces to $\kappa|\beta|^2\tau\ll1$. The probability that two photons reach the detectors within one collision, which is of order $[\kappa\tau(\langle\hat a^\dagger\hat a\rangle+|\beta|^2)]^2$, is then negligible, so that at most one photon reaches the detectors per collision, as assumed for the click operators above. The collision time $\tau$ is thus a discretization step of the model rather than a physical time resolution. A time interval resolved in an experiment, such as the sampling interval of the measurement record, is assumed to comprise many collisions while remaining much shorter than the time scales of the system dynamics. For a strong local oscillator, such an interval also contains many clicks, which is the regime of the diffusive SME. The homodyne signal is given by the difference of the click probabilities of the two detectors. For $\phi = 0$, the measured quadrature is $\hat{Q}=-i(\hat{a}-\hat{a}^\dagger)/2$, and the POVM elements are
	\begin{equation}
		\begin{aligned}
			\hat{E}_{Q_1}&=\hat{K}^\dagger_{Q_1}\hat{K}_{Q_1}=\frac{\kappa\tau}{2}\big[\hat{a}^\dagger\hat{a}-2|\beta|\hat{Q}+|\beta|^2\hat{\mathbb{I}}\big],\\
			\hat{E}_{Q_2}&=\hat{K}^\dagger_{Q_2}\hat{K}_{Q_2}=\frac{\kappa\tau}{2}\big[\hat{a}^\dagger\hat{a}+2|\beta|\hat{Q}+|\beta|^2\hat{\mathbb{I}}\big].
		\end{aligned}
	\end{equation}
	For $\phi = -\pi/2$, the measured quadrature is $\hat{I}=(\hat{a}+\hat{a}^\dagger)/2$, and
	\begin{equation}
		\begin{aligned}
			\hat{E}_{I_1}&=\hat{K}^\dagger_{I_1}\hat{K}_{I_1}=\frac{\kappa\tau}{2}\big[\hat{a}^\dagger\hat{a}-2|\beta|\hat{I}+|\beta|^2\hat{\mathbb{I}}\big],\\
			\hat{E}_{I_2}&=\hat{K}^\dagger_{I_2}\hat{K}_{I_2}=\frac{\kappa\tau}{2}\big[\hat{a}^\dagger\hat{a}+2|\beta|\hat{I}+|\beta|^2\hat{\mathbb{I}}\big].
		\end{aligned}
		\label{eq:I_phase}
	\end{equation}
	The homodyne signals are therefore
	\begin{equation}
		\begin{aligned}
			p_{Q_{\mathrm{homodyne}}} &= \langle\hat{E}_{Q_2}\rangle-\langle\hat{E}_{Q_1}\rangle=2\kappa\tau|\beta|\langle\hat{Q}\rangle,\\
			p_{I_{\mathrm{homodyne}}} &= \langle\hat{E}_{I_2}\rangle-\langle\hat{E}_{I_1}\rangle=2\kappa\tau|\beta|\langle\hat{I}\rangle.
		\end{aligned}
		\label{eq:hetero_signal}
	\end{equation}
	The local oscillator thus amplifies the quadrature signal by $|\beta|$, while its contribution to the click probabilities, proportional to $|\beta|^2$, sets the shot noise. In the notation of Sec.~\ref{sec:sme}, $\hat L=\sqrt{\kappa}\,\hat a$ and $\beta_k=-i\sqrt{\kappa}\,\beta$, so that the reference phase is $\phi_k=\phi-\pi/2$ because of the phase $-i$ acquired at the beam splitter. Up to sign and normalization, the quadratures $\hat Q$ and $\hat I$ selected by $\phi=0$ and $\phi=-\pi/2$ are therefore the quadratures $\hat x_{\pi/2}$ and $\hat x_0$ of Sec.~\ref{sec:sme}.
	
	\subsubsection{\label{sec:heterodyne}Heterodyne detection}
	
	Heterodyne detection in the double-homodyne, or eight-port, configuration of Fig.~\ref{fig:all-gates}(c) is the simplest composite scheme~\cite{WisemanMilburn2009,WisemanMilburn1993,ChiaWiseman2011}. The measured mode $\hat s$ and a vacuum mode $\hat b$ are mixed on a balanced beam splitter of the form of Eq.~\eqref{eq:eq_balanced_bm}, and each output enters a homodyne module, one measuring the I and the other the Q quadrature. The outputs of the vacuum beam splitter are
	\begin{equation}
		\begin{aligned}
			\hat f_2&=\frac{\hat s-i\hat b}{\sqrt2}\;\to\;\frac{\hat s}{\sqrt2}=\sqrt{\tfrac{\kappa\tau}{2}}\,\hat a,\\
			\hat f_1&=\frac{\hat b-i\hat s}{\sqrt2}\;\to\;\frac{-i\hat s}{\sqrt2}=-i\sqrt{\tfrac{\kappa\tau}{2}}\,\hat a,
		\end{aligned}
		\label{eq:f_proj}
	\end{equation}
	where the arrows denote projecting out the vacuum, using $\langle 0|\hat b|0\rangle=0$ and $\langle 0|0\rangle=1$. The only effect of this beam splitter is thus a factor $1/\sqrt2$ on the signal amplitude in each arm, together with a phase $-i$ in the $\hat f_1$ arm. Because the vacuum port carries no amplitude, this beam splitter adds no further dependence on $\kappa$ and $\tau$.
	
	The derivation of Eq.~\eqref{eq:homo_kraus} makes no assumption about the input mode of the homodyne module. It therefore holds for any input mode of the form $\sqrt{\kappa\tau}\,\hat f$, where $\hat f$ acts on the system, and for any local-oscillator eigenvalue $\beta_{\rm anc}'=\sqrt{\kappa\tau}\,\beta_{\rm anc}$, with $\hat a\to\hat f$ and $\beta\to\beta_{\rm anc}$. The heterodyne Kraus operators thus follow directly from the homodyne ones, with one homodyne module in each arm,
	\begin{equation}
		\begin{aligned}
			\hat K_{Q_j}&=\hat K_{R_j}\big|_{\hat a\to\hat a/\sqrt2}\\
			\hat K_{I_j}&=\hat K_{R_j}\big|_{\hat a\to-i\hat a/\sqrt2,\ \beta\to-\beta}
		\end{aligned}
		\qquad j=1,2,
		\label{eq:hetero_compose}
	\end{equation}
	with $\beta$ real and positive. The $\hat f_2$ arm uses the local oscillator $|\beta'\rangle$ (eigenvalue $+\beta'$), so that the phase $\phi = 0$ selects the Q quadrature. The $\hat f_1$ arm uses the local oscillator $|-\beta'\rangle$, i.e., $\beta_{\rm anc}=-\beta$ and $\phi = \pi$, instead of the phase $\phi = -\pi/2$ of Eq.~\eqref{eq:I_phase}. This choice compensates the phase $-\pi/2$ acquired at the first beam splitter [Eq.~\eqref{eq:f_proj}], so that this arm measures the I quadrature. Substituting Eq.~\eqref{eq:f_proj} into Eq.~\eqref{eq:hetero_compose} and pulling out the factor $1/\sqrt2$, we obtain the heterodyne click Kraus operators
	\begin{equation}
		\begin{aligned}
			\hat K_{Q_1}&
			=\sqrt{\tfrac{\kappa\tau}{4}}\,\big(\hat a-i\sqrt2\,\beta\hat{\mathbb I}\big),\\
			\hat K_{Q_2}&
			=\sqrt{\tfrac{\kappa\tau}{4}}\,\big(\sqrt2\,\beta\hat{\mathbb I}-i\hat a\big),\\
			\hat K_{I_1}&
			=\sqrt{\tfrac{\kappa\tau}{4}}\,\big(-i\hat a+i\sqrt2\,\beta\hat{\mathbb I}\big),\\
			\hat K_{I_2}
			&=\sqrt{\tfrac{\kappa\tau}{4}}\,\big(-\sqrt2\,\beta\hat{\mathbb I}-\hat a\big).
		\end{aligned}
		\label{eq:hetero_kraus}
	\end{equation}
	The entire heterodyne apparatus, comprising the vacuum beam splitter and two homodyne modules with their local oscillators and four detectors, has thus been folded into four click Kraus operators, one for each detector, that act only on the Hilbert space of the measured mode. With the two local oscillators, the total click probability of the four detectors per time step is $\kappa\tau(\langle\hat a^\dagger\hat a\rangle+2\beta^2)$, so the condition of Eq.~\eqref{eq:dt_condition} applies with $|\beta|^2$ replaced by $2\beta^2$.
	
	With $\hat Q=-i(\hat a-\hat a^\dagger)/2$, the POVM elements of the Q arm are
	\begin{equation}
		\begin{aligned}
			\hat E_{Q_1}&=\hat K_{Q_1}^\dagger\hat K_{Q_1}=\tfrac{\kappa\tau}{4}\big[\hat a^\dagger\hat a-2\sqrt2\,\beta\hat Q+2\beta^2\hat{\mathbb I}\big],\\
			\hat E_{Q_2}&=\hat K_{Q_2}^\dagger\hat K_{Q_2}=\tfrac{\kappa\tau}{4}\big[\hat a^\dagger\hat a+2\sqrt2\,\beta\hat Q+2\beta^2\hat{\mathbb I}\big],
		\end{aligned}
	\end{equation}
	and those of the I arm follow with $\hat Q\to\hat I=(\hat a+\hat a^\dagger)/2$. The heterodyne signals are therefore
	\begin{equation}
		\begin{aligned}
			p_{Q_{\mathrm{heterodyne}}} &= \langle\hat{E}_{Q_2}\rangle-\langle\hat{E}_{Q_1}\rangle=\sqrt2\kappa\tau\beta\langle\hat{Q}\rangle,\\
			p_{I_{\mathrm{heterodyne}}} &= \langle\hat{E}_{I_2}\rangle-\langle\hat{E}_{I_1}\rangle=\sqrt2\kappa\tau\beta\langle\hat{I}\rangle.
		\end{aligned}
		\label{eq:root2_scaling}
	\end{equation}
	Compared with Eq.~\eqref{eq:hetero_signal}, the signal per quadrature is reduced by a factor of $\sqrt2$, whereas the local-oscillator contribution to each click probability, and hence the shot noise, is unchanged. This is the 3-dB penalty of Sec.~\ref{sec:sme} for measuring two conjugate quadratures simultaneously, which provides a second consistency check. Here, its origin is explicit. As the substitution $\hat a\to\hat a/\sqrt2$ in Eq.~\eqref{eq:hetero_compose} shows, the first beam splitter sends only half of the signal power into each arm.
	
	Heterodyne detection illustrates a general property of our approach. For any detection chain composed of linear gates and fed by ancillas in eigenstates of their annihilation operators, the click operators of the final detectors can be propagated back to the measured mode. The resulting effective Kraus operators act on the Hilbert space of that mode alone and exactly reproduce the measurement statistics of the original click operators, which act on the joint Hilbert space of all modes. Moreover, the effective Kraus operators of a simpler scheme serve directly as a module for a more complex one, as the homodyne module does here for heterodyne detection and as the heterodyne module does for the gain--heterodyne readout of Sec.~\ref{sec:level5}.
	
	In Sec.~\ref{sec:level4}, we use the heterodyne module to describe quantum nondemolition (QND) dispersive readout of a superconducting qubit.

	\section{\label{sec:level4}Dispersive qubit readout}

		\subsection{\label{sec:disp_model}Dispersive readout model}

	We now apply the heterodyne module of Sec.~\ref{sec:heterodyne} to the dispersive readout of a superconducting qubit~\cite{Blais2004,Wallraff2004,Blais2021,Krantz2019}. When the qubit is strongly detuned from the readout resonator, the exchange of excitations between them is suppressed, and their interaction reduces to a qubit-state-dependent shift of the resonator frequency. With the Pauli operator $\hat\sigma_z=\ket{e}\bra{e}-\ket{g}\bra{g}$, where $\ket{e}$ and $\ket{g}$ are the excited and ground states of the qubit, and the annihilation operator $\hat a$ of the readout mode, the dispersive Hamiltonian of a qubit and a resonator driven at the frequency $\omega_d$ is~\cite{Blais2004,Blais2021,Krantz2019}
	\begin{equation}
		\begin{aligned}
			\hat H_{\mathrm{lab}}={}&\omega_r\hat a^\dagger\hat a+\frac{\omega_q}{2}\hat\sigma_z-\chi\hat\sigma_z\hat a^\dagger\hat a\\
			&+\Omega_D\left(\hat a\,e^{i\omega_d t}+\hat a^\dagger e^{-i\omega_d t}\right).
		\end{aligned}
		\label{eq:disp_lab}
	\end{equation}
	Here, $\omega_r$ and $\omega_q$ are the resonator and qubit frequencies, renormalized by their coupling, $\chi$ is the dispersive shift, and $\Omega_D$ is the amplitude of the coherent drive. For a two-level qubit coupled to the resonator with strength $g_{\mathrm{qr}}$ at the detuning $\Delta=\omega_q-\omega_r$, $\chi=-g_{\mathrm{qr}}^2/\Delta$, which is positive for a qubit below the resonator, and for a transmon, $\chi$ also depends on its anharmonicity~\cite{Koch2007,Blais2021}. The dispersive approximation requires the mean photon number in the resonator to remain well below the critical photon number $n_{\mathrm{crit}}=\Delta^2/4g_{\mathrm{qr}}^2$~\cite{Blais2004,Blais2021}, so the drive must be weak. In the frame rotating at $\omega_d$ for the resonator and at $\omega_q$ for the qubit, the Hamiltonian becomes time independent~\cite{Gambetta2008,Blais2021},
	\begin{equation}
		\hat H=\left(\omega_r-\omega_d-\chi\hat\sigma_z\right)\hat a^\dagger\hat a+\Omega_D(\hat a+\hat a^\dagger).
		\label{eq:full_ham}
	\end{equation}
	The resonator decays at the rate $\kappa$ into the output line that carries the signal to the detection chain, so the state $\hat\rho$ of qubit and resonator obeys the master equation [cf.\ Eq.~\eqref{eq:lindblad}]
	\begin{equation}
		\dot{\hat\rho}=-i[\hat H,\hat\rho]+\kappa\,\mathcal{D}[\hat a]\hat\rho ,
		\label{eq:disp_master}
	\end{equation}
	whose dissipator is unchanged by the transformation to the rotating frame. In our model, this dissipator is generated by the collisional beam splitter [Eqs.~\eqref{eq:U_BS_coll} and \eqref{eq:coll_lindblad}], and the heterodyne module unravels it into single measurement trajectories. We neglect intrinsic relaxation and dephasing of the qubit, losses of the resonator other than its coupling to the output line, and photon loss and classical noise in the detection chain.

	Because $\hat H$ commutes with $\hat\sigma_z$, the populations of $\ket{e}$ and $\ket{g}$ are conserved, and the readout is QND~\cite{Blais2004,Blais2021}. For a qubit in a basis state $\ket{j}$, $j\in\{e,g\}$, $\hat\sigma_z$ can be replaced by its eigenvalue, and the readout mode evolves under
	\begin{equation}
		\hat H_j=(\omega_j-\omega_d)\,\hat a^\dagger\hat a+\Omega_D(\hat a+\hat a^\dagger),\qquad \omega_{e/g}=\omega_r\mp\chi .
		\label{eq:H_j}
	\end{equation}
	The qubit thus acts only as a switch between two resonator frequencies separated by $2\chi$, while its own state is not changed. For each basis state, the readout mode is a driven, damped harmonic oscillator, whose field amplitude obeys $\tfrac{d}{dt}\langle\hat a\rangle=-[\kappa/2+i(\omega_j-\omega_d)]\langle\hat a\rangle-i\Omega_D$. Because this dynamics is linear and the resonator decays into the vacuum, an initially empty resonator remains in a coherent state $\ket{\alpha_j(t)}$, with
	\begin{equation}
		\begin{aligned}
			\alpha_j(t)&=\alpha_j\left\{1-e^{-[\kappa/2+i(\omega_j-\omega_d)]t}\right\},\\
			\alpha_j&=\frac{-i\,\Omega_D}{\kappa/2+i(\omega_j-\omega_d)} .
		\end{aligned}
		\label{eq:alpha_ss}
	\end{equation}
	The steady-state amplitude $\alpha_j$ is the Lorentzian response of a resonator with frequency $\omega_j$ to the drive at $\omega_d$, and it is approached on the ring-up time scale $2/\kappa$. Since $\omega_e\neq\omega_g$, the two basis states of the qubit lead to two different steady states of the field, the pointer states $\ket{\alpha_e}$ and $\ket{\alpha_g}$~\cite{Gambetta2008,Blais2021}. For a drive midway between the two resonator frequencies, $\omega_d=\omega_r$,
	\begin{equation}
		\alpha_{e/g}=\frac{\Omega_D\,(\pm\chi-i\kappa/2)}{\kappa^2/4+\chi^2} .
		\label{eq:alpha_sym}
	\end{equation}
	The two pointer states then contain the same mean number of photons, $|\alpha_j|^2=\Omega_D^2/(\kappa^2/4+\chi^2)$, and differ only in phase, which is shifted by $\pm\arctan(2\chi/\kappa)$ relative to the response of a resonator at $\omega_r$~\cite{Blais2021}. Their difference, $\alpha_e-\alpha_g=2\chi\Omega_D/(\kappa^2/4+\chi^2)$, is real, so the qubit state is encoded in the I quadrature of the field, whereas the Q quadrature, $\mathrm{Im}\,\alpha_j=-\kappa\Omega_D/[2(\kappa^2/4+\chi^2)]$, is the same for both states. A phase-sensitive measurement is therefore required, since photon counting cannot distinguish the two states (Sec.~\ref{sec:sme}). The readout mode emits the field $\hat s=\sqrt{\kappa\tau}\,\hat a$ into the detection chain in each collision, and by Eq.~\eqref{eq:root2_scaling}, the mean click differences of the I and Q arms of the heterodyne module are proportional to $\mathrm{Re}\langle\hat a\rangle$ and $\mathrm{Im}\langle\hat a\rangle$. In a single run of the readout, or shot, the clicks of each detector are counted over an integration time $t_{\rm int}$. The outcome of the shot is a point in the IQ plane, whose coordinates are the differences between the click counts of the two detectors of the I arm and of the Q arm, taken with the signs of Eq.~\eqref{eq:root2_scaling}. For a qubit in $\ket{j}$, the mean outcome is thus proportional to $\int_0^{t_{\rm int}}\alpha_j(t)\,dt$, which is dominated by $\alpha_j t_{\rm int}$ for $t_{\rm int}\gg2/\kappa$.

	\begin{figure*}
		\centering
		\includegraphics[width=0.48\textwidth]{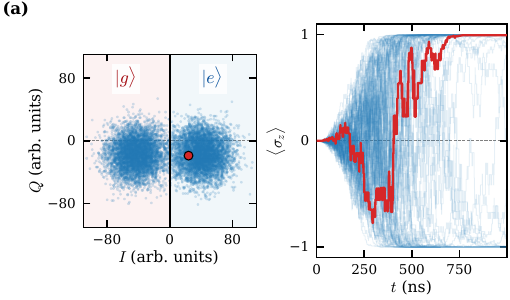}\hfill
		\includegraphics[width=0.48\textwidth]{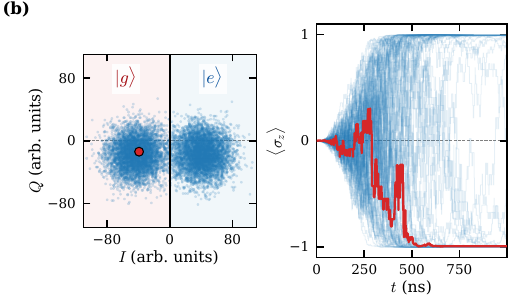}

		\vspace{0.8em}

		\includegraphics[width=0.48\textwidth]{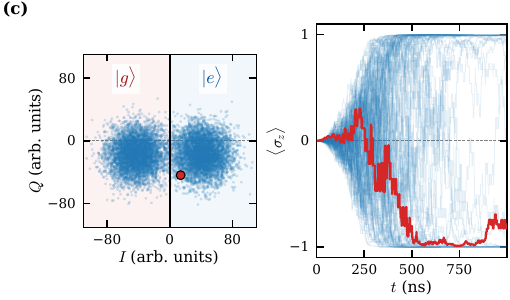}\hfill
		\includegraphics[width=0.48\textwidth]{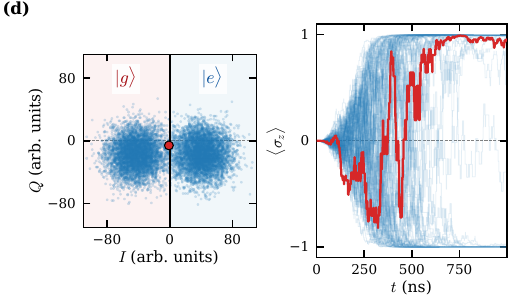}

		\caption{Single-shot trajectories of the QND heterodyne readout for a qubit initially in $\ket{+}=(\ket{e}+\ket{g})/\sqrt2$, illustrating the four possible outcomes of state assignment. In each panel, the left plot shows the outcomes of all shots in the IQ plane (blue), which are the same in all panels, together with one highlighted shot (red). The right plot shows the trajectory of $\langle\hat\sigma_z\rangle$ for the highlighted shot (red) over the trajectories of other shots (light blue). A shot is assigned to $\ket{e}$ if it lies to the right of the decision boundary (vertical black line) and to $\ket{g}$ if it lies to its left, and the assignment is compared with the final value of $\langle\hat\sigma_z\rangle$ of its trajectory. (a)~Correct assignment, with the shot to the right of the boundary and the qubit collapsing onto $\ket{e}$ ($\langle\hat\sigma_z\rangle\to1$).
			(b)~Correct assignment, with the shot to the left of the boundary and the qubit collapsing onto $\ket{g}$ ($\langle\hat\sigma_z\rangle\to-1$). (c)~Wrong assignment, with the shot to the right of the boundary but the qubit collapsing onto $\ket{g}$ ($\langle\hat\sigma_z\rangle\to-1$). (d)~Wrong assignment, with the shot just to the left of the boundary but the qubit collapsing onto $\ket{e}$ ($\langle\hat\sigma_z\rangle\to1$). The parameters are $\kappa/2\pi=1$~MHz, $\chi/2\pi=1.5$~MHz, $\Omega_D/2\pi=2$~MHz, $|\beta|^2=400$, and $t_{\rm int}=1000$~ns.}
		\label{fig:four_panels}
	\end{figure*}

	\subsection{\label{sec:assign}State collapse and finite-time assignment error}

	Even for an ideal detection chain, the outcome of a shot is random. The clicks of each detector are Poissonian, and for a strong local oscillator, their fluctuations reflect the vacuum fluctuations of the emitted field, which the local oscillator amplifies together with the signal, as well as the vacuum that enters through the unused port of the first beam splitter of the heterodyne module (Sec.~\ref{sec:heterodyne})~\cite{WisemanMilburn2009,Clerk2010}. This noise is of quantum-mechanical origin and cannot be removed by better detectors. Its consequences are simplest for a qubit in a basis state $\ket{j}$, for which it suffices to evolve the readout mode under $\hat H_j$ of Eq.~\eqref{eq:H_j}, with the heterodyne Kraus operators acting on it. Repeated shots then produce a distribution of outcomes, a blob in the IQ plane centered at the mean outcome. By Eq.~\eqref{eq:alpha_sym}, the blobs of $\ket{e}$ and $\ket{g}$ have the same Q coordinate and are separated along the I quadrature, with $\ket{e}$ corresponding to the right blob. Since the clicks in different collisions are independent, the distance between the two blob centers grows linearly with $t_{\rm int}$, whereas the widths of the blobs grow only as $\sqrt{t_{\rm int}}$. The two blobs would therefore separate completely only for $t_{\rm int}\to\infty$. Any measurement, however, takes a finite time, after which the blobs still overlap.

	To assign a state to a shot, we use the decision boundary midway between the two blob centers and perpendicular to the line that joins them. A shot is assigned to $\ket{e}$ if it lies to the right of this boundary and to $\ket{g}$ otherwise. Since the two blob centers differ only in I, this rule compares the integrated I quadrature of the shot with its value midway between the two centers. Because the blobs overlap, a shot from $\ket{e}$ can land on the side of the decision boundary that belongs to $\ket{g}$, and vice versa. The probability of such a misassignment is the assignment error. It is set by the separation of the two pointer states relative to the quantum noise accumulated during $t_{\rm int}$. In the steady state, the distance between the blob centers relative to their widths grows linearly with the separation $|\alpha_e-\alpha_g|$ of the two pointer states and with $\sqrt{t_{\rm int}}$, so the assignment error decreases with both. The ring-up of the resonator increases the assignment error, because the pointer states are barely separated during the ring-up, so that the record accumulates noise but little information about the qubit.

	For a basis state, the qubit keeps its initial state, and the readout only has to decide whether the field has settled at $\alpha_e$ or $\alpha_g$. For a superposition $\cos(\theta/2)\ket{e}+e^{i\phi}\sin(\theta/2)\ket{g}$, such as the equal superposition $\ket{+}=(\ket{e}+\ket{g})/\sqrt2$ with $\theta=\pi/2$ and $\phi=0$, the field instead settles at one of the two values, with probabilities given by the initial state of the qubit. Because $\hat H$ commutes with $\hat\sigma_z$ and the heterodyne Kraus operators act only on the readout mode, each branch evolves as for the corresponding basis state. Along a single trajectory, the joint state of qubit and readout mode therefore remains of the form~\cite{Gambetta2008,Korotkov2016,Blais2021}
	\begin{equation}
		\begin{aligned}
			\ket{\Psi(t)}\propto{}&\cos\frac{\theta}{2}\,\lambda_e(t)\ket{e}\ket{\alpha_e(t)}\\
			&+e^{i\phi}\sin\frac{\theta}{2}\,\lambda_g(t)\ket{g}\ket{\alpha_g(t)},
		\end{aligned}
		\label{eq:cond_state}
	\end{equation}
	where the amplitudes $\alpha_j(t)$ of Eq.~\eqref{eq:alpha_ss} evolve deterministically, and only the complex weights $\lambda_j(t)$, with $\lambda_j(0)=1$, depend on the measurement record. A click only multiplies a coherent state by a number, because the Kraus operators of Eq.~\eqref{eq:hetero_kraus} are linear combinations of $\hat a$ and the identity, whereas the evolution between clicks keeps the state coherent with the damped amplitude $\alpha_j(t)$. The weight $|\lambda_j(t)|^2$ is proportional to the probability of the record for a qubit in $\ket{j}$, so the conditional populations are the initial populations weighted by these probabilities and renormalized~\cite{Korotkov2016}. For $\omega_d=\omega_r$, $\alpha_g(t)=-\alpha_e^*(t)$, so only the clicks of the I arm change the populations, whereas the clicks of the Q arm only kick the relative phase of the two branches~\cite{Korotkov2016,Hatridge2013}. Averaged over the records, the populations keep their initial values $\cos^2(\theta/2)$ and $\sin^2(\theta/2)$. As the two pointer states become distinguishable, one weight dominates, and the qubit collapses to $\ket{e}$ or $\ket{g}$, and the field to $\alpha_e$ or $\alpha_g$, with these Born-rule probabilities~\cite{Hatridge2013,Murch2013}. Because the two branches of Eq.~\eqref{eq:cond_state} are orthogonal, the probability of a record for the superposition is the sum of its probabilities for $\ket{e}$ and $\ket{g}$, weighted by the initial populations. The shots of a superposition therefore fall into the two blobs of the basis states, with probabilities given by the initial populations. Along a trajectory, the magnitude of the qubit coherence is $\sqrt{P_eP_g}\,|\langle\alpha_g(t)|\alpha_e(t)\rangle|$, where $P_j$ are the conditional populations, so it decreases both through the collapse and through the entanglement of the qubit with the intracavity field as the pointer states separate. Without conditioning on the record, the qubit coherence decays at the measurement-induced dephasing rate $\Gamma_m=\kappa|\alpha_e-\alpha_g|^2/2$ in the steady state~\cite{Gambetta2006,Gambetta2008,Blais2021}. The separation $|\alpha_e-\alpha_g|$ that sets the distance between the two blobs thus also sets the rate of the collapse.

	To follow this collapse, we evolve the qubit and the readout mode jointly under Eq.~\eqref{eq:disp_master}, unraveled by the heterodyne Kraus operators. Figure~\ref{fig:four_panels} shows the result for a qubit initially in $\ket{+}$. In its IQ panels, the shots form two blobs of about equal weight, as expected for $\ket{+}$, and these are the blobs of $\ket{e}$ and $\ket{g}$. Both blobs lie at the same negative Q, which reflects the common imaginary part of $\alpha_e$ and $\alpha_g$, and they are separated along I, with $\ket{e}$ on the right. At $t_{\rm int}=1000$~ns, the two blobs are clearly distinguishable, but their tails still overlap near the decision boundary. The trajectories of $\langle\hat\sigma_z\rangle$ in Fig.~\ref{fig:four_panels} show how the collapse happens. All trajectories start at $\langle\hat\sigma_z\rangle=0$ and stay close to zero at early times, while the resonator rings up on the time scale $2/\kappa\approx320$~ns and the pointer states are barely separated. They then spread, and most of them reach $+1$ or $-1$ within the integration time, i.e., the qubit collapses onto $\ket{e}$ or $\ket{g}$. Together with the IQ panels, these trajectories show how a projective measurement of $\hat\sigma_z$ is realized in practice by a heterodyne readout chain. In textbook quantum mechanics, a projective measurement is an instantaneous collapse onto an eigenstate, with probabilities given by the Born rule. Here, the same outcome emerges gradually from the stochastic clicks of the detectors, each of which changes the state of the qubit only slightly, in accordance with the theory of continuous measurement~\cite{Brun2002,JacobsSteck2006,Gambetta2008,Korotkov2016} and with experiments that tracked single trajectories of superconducting qubits~\cite{Murch2013,Hatridge2013}. In the spirit of Ref.~\cite{Korotkov2016}, which applies the textbook collapse postulate to the pieces of the microwave field that leak out of the resonator, the collision model makes this emergence explicit at the level of individual photocounts and thereby connects the operation of a realistic readout chain with the foundational notion of a projective measurement.

	For a superposition, the prepared state does not determine the correct outcome. To quantify the assignment error for an arbitrary initial state, we therefore compare, for each shot, the blob in which it lands with the final state of its trajectory~\cite{Gambetta2008,Hatridge2013,Murch2013}. The blob is determined by the side of the decision boundary on which the shot lands, and the final state by the sign of $\langle\hat\sigma_z\rangle$ at the end of the trajectory. A shot is wrongly assigned if the two disagree. For an initial $\hat\sigma_z$ eigenstate, this definition coincides with the usual one, which compares the assigned state with the prepared state. Figure~\ref{fig:four_panels} shows one shot for each of the four possible combinations. In Figs.~\ref{fig:four_panels}(a) and \ref{fig:four_panels}(b), the shot lies on the side of the decision boundary that belongs to the final state of the qubit, and the assignment is correct. In Fig.~\ref{fig:four_panels}(b), the shot lies close to the center of the $\ket{g}$ blob, and after a small excursion toward $\ket{e}$, the qubit collapses onto $\ket{g}$ by about 500~ns. In Fig.~\ref{fig:four_panels}(a), the shot lies between the decision boundary and the center of the $\ket{e}$ blob, and the collapse is far from monotonic. The record first favors $\ket{g}$, so that $\langle\hat\sigma_z\rangle$ drops to about $-0.75$ near 300~ns, and only later favors $\ket{e}$, onto which the qubit collapses after about 700~ns. In Figs.~\ref{fig:four_panels}(c) and \ref{fig:four_panels}(d), the shot lies close to the decision boundary, and the qubit collapses onto the state of the opposite blob. In Fig.~\ref{fig:four_panels}(c), $\langle\hat\sigma_z\rangle$ first rises slightly toward $\ket{e}$ and then reaches $-1$ near the middle of the integration window, before it returns to about $-0.8$ near the end of the window. In Fig.~\ref{fig:four_panels}(d), the shot lies almost on the decision boundary, on the side of $\ket{g}$. The record first favors $\ket{g}$, so that $\langle\hat\sigma_z\rangle$ drops to about $-0.8$ near 300~ns and swings back and forth near 400~ns, and the qubit collapses onto $\ket{e}$ only in the second half of the integration window. These two shots show how assignment errors arise at a finite integration time. Near the decision boundary, the integrated record does not yet separate the two outcomes, so the blob in which a shot lands can disagree with the state onto which the qubit collapses.

\begin{figure}
		\centering
		\includegraphics[width=\columnwidth]{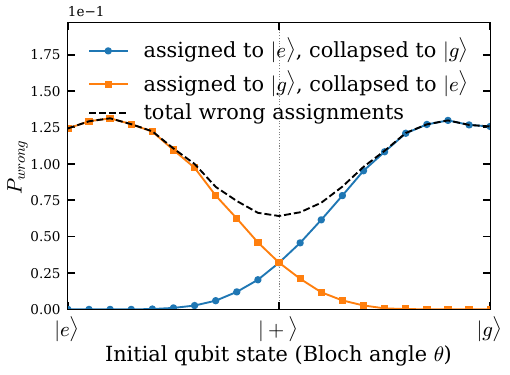}
		\caption{Probability $P_{\rm wrong}$ of a wrong assignment as a function of the initial qubit state $\cos(\theta/2)\ket{e}+\sin(\theta/2)\ket{g}$ along the Bloch-sphere path from $\ket{e}$ through $\ket{+}$ to $\ket{g}$ (Bloch angle $\theta$), estimated from $n_{\rm traj}=400\,000$ trajectories per point. Blue circles (orange squares) show shots that are assigned to $\ket{e}$ ($\ket{g}$), i.e., lie to the right (left) of the decision boundary, but whose qubit collapses onto $\ket{g}$ ($\ket{e}$). The dashed line shows the total. The parameters are $\kappa/2\pi=1$~MHz, $\chi/2\pi=1.5$~MHz, $\Omega_D/2\pi=2$~MHz, $|\beta|^2=400$, and $t_{\rm int}=1000$~ns.}
		\label{fig:assign_err}
	\end{figure}

	These results raise the question whether the finite-time assignment error is the same for all initial states of the qubit. Since the qubit does not decohere and the detection chain is ideal, the assignment error is set only by the quantum noise of the readout chain and the finite integration time, and one might expect it to be independent of the initial state. Figure~\ref{fig:assign_err} shows the probability of a wrong assignment, i.e., the fraction of misassigned shots, for initial states along the path from $\ket{e}$ through $\ket{+}$ to $\ket{g}$ on the Bloch sphere. The assignment error clearly depends on the initial state. It is largest near the $\hat\sigma_z$ eigenstates and about half as large for the equal superposition $\ket{+}$. Moreover, near $\ket{e}$, almost all misassigned shots land in the $\ket{g}$ blob while the qubit remains close to $\ket{e}$, and vice versa near $\ket{g}$.

	This dependence follows from the conditional state of Eq.~\eqref{eq:cond_state}, in which the record changes the qubit populations only by reweighting the initial populations. The sign of the final $\langle\hat\sigma_z\rangle$ is thus the more probable outcome given the initial populations and the full record, in which each instant is weighted by the separation of the two pointer states~\cite{Gambetta2008,Walter2017}. Our assignment rule, in contrast, weights the record uniformly and ignores the initial populations. For an initial $\hat\sigma_z$ eigenstate, the conditional state does not change, so every shot that lands in the wrong blob is misassigned, and the error equals the overlap of the two blobs. For $\ket{+}$, the same record that determines the blob also drives the collapse~\cite{Hatridge2013,Murch2013}, and the two rules disagree only for shots near the decision boundary, such as those in Figs.~\ref{fig:four_panels}(c) and \ref{fig:four_panels}(d). They disagree mainly because the uniform weighting also counts the ring-up of the readout mode, during which the pointer states are barely separated. In both of these shots, $\langle\hat\sigma_z\rangle$ first moves toward the state of the blob in which the shot lands, while the pointer states are still separating, and only later toward the state onto which the qubit collapses. Without such transients, uniform weighting would be optimal~\cite{Gambetta2007}. For intermediate states, the unequal populations shift the decision boundary of the conditional state away from the midpoint. This produces the smooth dependence in Fig.~\ref{fig:assign_err} and explains why, near each eigenstate, almost all misassigned shots are of one type. The lower error for $\ket{+}$ therefore does not indicate a better readout but reflects the correlation between the record and the collapse.

	The finite-time assignment error discussed in this subsection arises under ideal conditions. In our model, the detection chain loses no photons and adds no classical noise, such as the noise that amplifiers add beyond the quantum limit. Moreover, neither the qubit nor the resonator undergoes dissipation or decoherence, apart from the decay of the resonator into the output line. In practice, all of these imperfections are present. Classical noise broadens the blobs, photon loss reduces their separation relative to their widths, and relaxation of the qubit during the readout changes the state that is being measured, so the assignment error of an actual readout chain is larger. In contrast, the assignment error found here is fundamental. It is present at any finite integration time even in an ideal readout chain, and it originates from the vacuum fluctuations of the electromagnetic field and from the stochastic collapse of the qubit state. A weighting of the record that is adapted to the ring-up of the readout mode can reduce this error but cannot remove it.

	Throughout this subsection, the local oscillator is strong, with $|\beta|^2 \gg 1$, so many clicks accumulate within $t_{\rm int}$ and the blobs are close to Gaussian. Because the collision model retains the individual photocounts, it also describes a weak local oscillator, for which the blobs reflect the discreteness of the counts, as we discuss next.


	\begin{figure*}
		\centering
		\setlength{\availW}{0.97\textwidth}   

		\sbox{\lbox}{\vbox{%
				\hbox{\includegraphics[width=5cm]{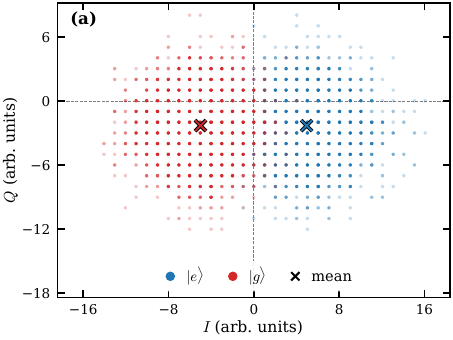}}\nointerlineskip
				\hbox{\includegraphics[width=5cm]{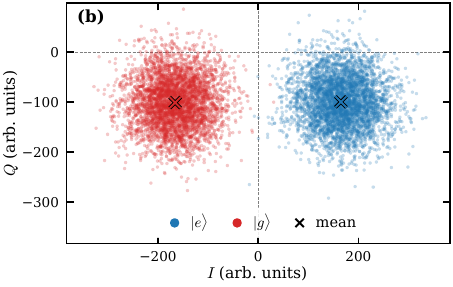}}}}
		\sbox{\rbox}{\hbox{\includegraphics[width=5cm]{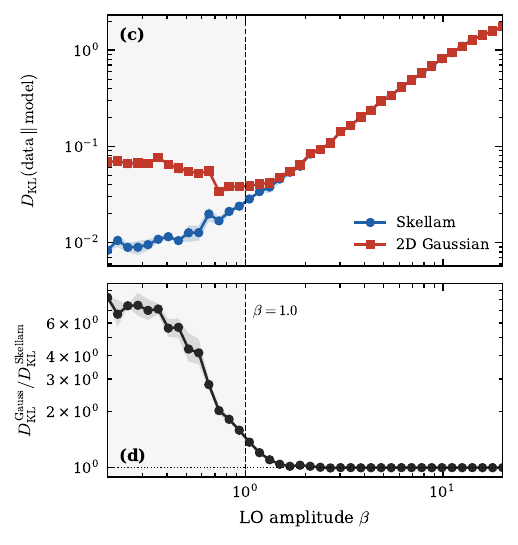}}}
		\setlength{\hL}{\dimexpr\ht\lbox+\dp\lbox\relax}
		\setlength{\hR}{\dimexpr\ht\rbox+\dp\rbox\relax}

		\setlength{\wL}{\dimexpr\availW*\numexpr\hR\relax/\numexpr\hL+\hR\relax\relax}
		\setlength{\wR}{\dimexpr\availW-\wL\relax}

		\noindent
		\vbox{%
			\hbox{\includegraphics[width=\wL]{fig8a.pdf}}\nointerlineskip
			\hbox{\includegraphics[width=\wL]{fig8b.pdf}}}%
		\hfill
		\vbox{\hbox{\includegraphics[width=\wR]{fig7.pdf}}}

		\caption{Skellam-to-Gaussian crossover of the single-shot outcomes as a function of the local-oscillator amplitude $\beta$~\cite{Skellam1946}. (a)~$\beta = 0.5$. (b)~$\beta = 20$. In (a) and (b), the outcomes for $\ket{e}$ (blue) and $\ket{g}$ (red) are shown in the IQ plane, and crosses mark their means. (c)~Kullback--Leibler (KL) divergences of the 2D Skellam and 2D Gaussian models from the simulated distribution as functions of $\beta$. (d)~Ratio of the two KL divergences~\cite{Skellam1946, 10.1214/aoms/1177729694}. The dashed line marks $\beta=1$, shortly above which the ratio approaches unity, and the shading marks the region $\beta<1$. The ratio drops most steeply around $\beta_c\approx0.7$. The parameters are $\kappa/2\pi=1$~MHz, $\chi/2\pi=1.5$~MHz, $\Omega_D/2\pi=2$~MHz, and $t_{\rm int}=1000$~ns.}
		\label{fig:IQ_transition}
	\end{figure*}

	\subsection{\label{sec:skellam}Skellam-to-Gaussian crossover}

	The discreteness of the photocounts becomes visible for a weak local oscillator, i.e., for small $|\beta|^2$, where $\kappa|\beta|^2$ is the photon flux of each of the two local oscillators of the heterodyne module [Eq.~\eqref{eq:beta_scaling}]. Since each of its four detectors [Figs.~\ref{fig:all-gates}(b) and \ref{fig:all-gates}(c)] registers at most one photon per collision, the count of each detector accumulated over the integration time is Poisson distributed. Each coordinate of the outcome, the difference between the counts of the two detectors of one arm, therefore follows a Skellam distribution~\cite{Skellam1946}. For small $|\beta|^2$, only a few clicks are registered, and the outcomes are visibly discrete and non-Gaussian [Fig.~\ref{fig:IQ_transition}(a)]. For $|\beta|^2\gg1$, where typical superconducting-qubit readout takes place, the counts are large, and by the central limit theorem, each blob approaches a Gaussian distribution [Fig.~\ref{fig:IQ_transition}(b)].

	Deviations from Gaussian statistics at weak local oscillators are known for balanced homodyne detection~\cite{VogelGrabow1993,TycSanders2004}. Here, they emerge directly from the click operators of the heterodyne module, without additional modeling, whereas the diffusive SME of Sec.~\ref{sec:sme} yields Gaussian statistics by construction.

	We quantify this crossover by the Kullback--Leibler (KL) divergence between the simulated distribution of outcomes and each of the two models~\cite{10.1214/aoms/1177729694}. For a measurement of $N = 4096$ shots, we determine the mean and the standard deviation of a two-dimensional (2D) Gaussian model and the Poisson means of a 2D Skellam model from the raw data and define the KL divergence of each model as a function of $\beta$ as
	\begin{equation}
		\begin{split}
			D_{\mathrm{KL}}(\beta) = {}&-\sum_{n} P_\mathrm{Num}(\mathbf{x}_n|\beta)\ln\frac{P_\mathrm{G/S}(\mathbf{x}_n|\beta)}{P_\mathrm{Num}(\mathbf{x}_n|\beta)},
			\label{eq:KL_eq}
		\end{split}
	\end{equation}
	where $P_\mathrm{Num}$ and $P_\mathrm{G/S}$ denote the probabilities of the outcome $\mathbf{x}_n$ in the numerical simulation and in the Gaussian or Skellam model, respectively, and the sum runs over all possible outcomes. Since each coordinate of the outcome is a difference of two counts, the outcomes $\mathbf{x}_n$ are pairs of integers. The KL divergence vanishes if a model reproduces the simulated distribution exactly, and a lower value indicates a better fit.

	Figures~\ref{fig:IQ_transition}(c) and \ref{fig:IQ_transition}(d) show the KL divergences of the two models and their ratio as functions of the local-oscillator amplitude $\beta$. For small $\beta$, the Skellam model reproduces the shots markedly better than the Gaussian model. The ratio of the two divergences drops sharply around $\beta_c\approx0.7$ and approaches unity shortly above $\beta=1$, where, for the present parameters, enough clicks accumulate within the integration time for the central limit theorem to apply. For larger $\beta$, both models reproduce the shots equally well, as expected, since the Skellam distribution becomes equivalent to the Gaussian distribution in the limit $\beta\rightarrow\infty$. Both divergences nevertheless increase with $\beta$ in this regime. This increase is caused by the finite number of shots and does not indicate that either model fails. The widths of the blobs, measured in units of single counts, grow in proportion to $\beta$. The same $N=4096$ shots are therefore distributed over more and more possible outcomes, and each outcome is registered by fewer shots. The simulated probabilities $P_\mathrm{Num}$ in Eq.~\eqref{eq:KL_eq} then become less accurate estimates of the true distribution, which increases the KL divergences of both models alike and leaves their ratio at unity.

	So far, the detection chain has been ideal, and the emitted field has reached the detectors without amplification. In Sec.~\ref{sec:level5}, we include phase-preserving gain by replacing only the collision gate, reuse the heterodyne module unchanged, and analyze how the gain acts on the readout mode and on the distinguishability of the two blobs.
	
	\section{\label{sec:level5}Phase-preserving gain in the collision}
	
	Amplification is a key step in the readout of weak microwave fields~\cite{Clerk2010,Castellanos_Beltran_2008}, and phase-preserving parametric amplifiers are routinely used in superconducting-qubit readout~\cite{RoyDevoret2016,Bergeal2010,Macklin2015}. A phase-preserving amplifier transforms a mode $\hat x$ into $\cosh r\,\hat x+i\sinh r\,\hat v^\dagger$, with power gain $\cosh^2 r$. The idler mode $\hat v$ is required to preserve the bosonic commutation relation, and with the idler in the vacuum, the amplifier adds the minimum noise allowed by quantum mechanics~\cite{Caves1982,Clerk2010,Caves_2012}.
	
	Including such an amplifier in the collision model requires care. The mean number of photons that reach the detectors in one collision must be of order $\tau$, so that at most one photon reaches them per collision (Sec.~\ref{sec:homodyne}). For the local oscillator, this requirement underlies the scaling of Eq.~\eqref{eq:beta_scaling}. An amplifier that acts on every bin after the collision violates it, because even for vacuum inputs, it adds $\sinh^2 r$ photons to each bin, independently of $\tau$, so that the click rate diverges as $\tau\to0$. Such an amplifier acts independently on bins of vanishing duration and therefore has an infinite bandwidth. A physical amplifier has a finite bandwidth, and its description requires an additional mode cascaded with the readout mode~\cite{Gardiner1993,Carmichael1993PRL,GoughJames2009,Combes2017}, which we leave for future work. Here, we instead include the gain in the collision itself. In the modular framework, this requires replacing only the collision gate, whereas the heterodyne module of Sec.~\ref{sec:heterodyne} is reused unchanged.
	
	\subsection{\label{sec:gain_collision}Collision with gain}
	
	In each collision, the readout mode $\hat a$ interacts with a fresh readout bin $\hat t$ in the vacuum and a fresh idler bin $\hat v$ in the coherent state $\ket{\nu}$ through the three-mode unitary [Fig.~\ref{fig:all-gates}(d)]
	\begin{equation}
		\begin{aligned}
			U_{\mathrm{BG}}(g\tau,r)&=e^{-i\tau g\hat H_{\mathrm{BG}}},\\
			\hat H_{\mathrm{BG}}&=\cosh r\,\big(\hat a^\dagger\hat t+\hat a\hat t^\dagger\big)
			+i\sinh r\,\big(\hat t^\dagger\hat v^\dagger-\hat t\hat v\big),
		\end{aligned}
		\label{eq:U_BG}
	\end{equation}
	with the same coupling $g=\sqrt{\kappa/\tau}$ as in Eq.~\eqref{eq:sqrt_scaling}. We vary the gain parameter $r$ at fixed $g$, so that $\kappa$ remains the decay rate in the absence of gain. The first term of $\hat H_{\mathrm{BG}}$ is the beam-splitter, or exchange, interaction $\hat H_{\mathrm{BS}}$ of Eq.~\eqref{eq:U_BS_coll}, weighted by $\cosh r$, which converts excitations of the readout mode into excitations of the readout bin. The second term, weighted by $\sinh r$, is a two-mode-squeezing interaction, which creates photon pairs in the readout and idler bins. For $r=0$, $U_{\mathrm{BG}}$ reduces to the collisional beam splitter $U_{\mathrm{BS}}(g\tau)$. Between microwave modes, beam-splitter and two-mode-squeezing interactions are realized by parametric couplers pumped at the difference and at the sum of the mode frequencies, where they are known as frequency conversion and parametric amplification, respectively~\cite{Bergeal2010,Abdo2013PRL}. Both processes can be activated simultaneously, as exploited in nonreciprocal amplifiers and circulators~\cite{MetelmannClerk2015,Sliwa2015,Lecocq2017}. Equation~\eqref{eq:U_BG} describes an idealized, memoryless coupler of this kind between the readout mode, its output line, which carries the readout bins, and an idler line, which carries the idler bins.
	
	The action of $U_{\mathrm{BG}}$ follows from a simple observation. The operator $\cosh r\,\hat a+i\sinh r\,\hat v^\dagger$ obeys the bosonic commutation relation, since $\cosh^2 r-\sinh^2 r=1$, and $\hat H_{\mathrm{BG}}$ is the exchange interaction between $\hat t$ and this operator. In analogy with Eq.~\eqref{eq:operator_scaling}, the readout bin therefore leaves the collision, exactly, in the mode
	\begin{equation}
		\begin{aligned}
			\hat s={}&\hat t\cos g\tau\\
			&-i\big(\cosh r\,\hat a+i\sinh r\,\hat v^\dagger\big)\sin g\tau .
		\end{aligned}
		\label{eq:gain_bin_exact}
	\end{equation}
	Projecting out the vacuum bin $\hat t$, expanding to lowest order with $g\tau=\sqrt{\kappa\tau}$, and omitting the phase $-i$ as in Sec.~\ref{sec:coll_bs}, we obtain
	\begin{equation}
		\hat s=\sqrt{\kappa\tau}\,\big(\cosh r\,\hat a+i\sinh r\,\hat v^\dagger\big).
		\label{eq:gain_eq}
	\end{equation}
	Relative to the field $\sqrt{\kappa\tau}\,\hat a$ emitted in the absence of gain, the field that enters the readout bin is thus amplified in a phase-preserving manner, with power gain $\cosh^2 r$, and the idler bin supplies the added noise. The readout mode itself is not amplified, because only the field that enters the readout bin is amplified. 
	Both terms of Eq.~\eqref{eq:gain_eq} carry the factor $\sqrt{\kappa\tau}$, so the mean photon number of the readout bin after the collision is of order $\kappa\tau$. This holds for a coherent idler with any fixed amplitude $\nu$ per bin. Unlike the local-oscillator amplitude $\beta'$ of Eq.~\eqref{eq:beta_scaling}, $\nu$ therefore need not scale with $\tau$, and we keep it fixed as $\tau\to0$. 
	
	\subsection{\label{sec:gain_kraus}Gain--heterodyne Kraus operators}
	
	The readout bin is measured by the heterodyne module of Sec.~\ref{sec:heterodyne}, and the idler bin is discarded. Equation~\eqref{eq:gain_eq} differs from $\hat s=\sqrt{\kappa\tau}\,\hat a$ only by the substitution $\hat a\to\cosh r\,\hat a+i\sinh r\,\hat v^\dagger$. The vacuum port and the local oscillators of the heterodyne module are independent of the idler and are projected out exactly as in Sec.~\ref{sec:heterodyne}. The heterodyne Kraus operators of Eq.~\eqref{eq:hetero_kraus} can therefore be reused with this substitution, which gives the effective Kraus operators for the clicks of the four detectors,
	\begin{equation}
		\hat K_{\ell}=\sqrt{\tfrac{\kappa\tau}{4}}\,\big[x_\ell\big(\cosh r\,\hat a+i\sinh r\,\hat v^\dagger\big)+y_\ell\hat{\mathbb I}\big],
		\label{eq:Kxy}
	\end{equation}
	where $\ell\in\{Q_1,Q_2,I_1,I_2\}$ labels the detector, and the coefficients follow from Eq.~\eqref{eq:hetero_kraus},
	\begin{equation}
		\begin{gathered}
			x_{Q_1}=1,\quad x_{Q_2}=-i,\quad x_{I_1}=-i,\quad x_{I_2}=-1,\\
			y_{Q_1}=-i\sqrt2\beta,\quad y_{Q_2}=\sqrt2\beta,\\
			y_{I_1}=i\sqrt2\beta,\quad y_{I_2}=-\sqrt2\beta .
		\end{gathered}
		\label{eq:xy}
	\end{equation}
	
	The operators of Eq.~\eqref{eq:Kxy} still act on the idler bin. Unlike the local oscillator, the idler cannot be projected out as in Eq.~\eqref{eq:gen_uni_eq}, because $\hat K_{\ell}$ contains $\hat v^\dagger$, of which $\ket{\nu}$ is not an eigenstate. Physically, the two-mode-squeezing term creates photons in pairs, so a click can be accompanied by an additional photon in the idler bin, which then ends in a state orthogonal to its initial state. We therefore resolve the final state of the discarded idler bin in an orthonormal basis, following Eq.~\eqref{eq:coll_map}. With the displacement operator $\hat D(\nu)$, which satisfies $\hat D(\nu)\ket{0}=\ket{\nu}$ and $\hat D^\dagger(\nu)\,\hat v^\dagger\hat D(\nu)=\hat v^\dagger+\nu^*$, we have $\hat v^\dagger\ket{\nu}=\nu^*\ket{\nu}+\hat D(\nu)\ket{1}$, and hence
	\begin{equation}
		\hat K_{\ell}\big(\hat{\mathbb I}\otimes\ket{\nu}\big)
		=\hat K_{\ell;s}\otimes\ket{\nu}+\hat K_{\ell;n}\otimes\hat D(\nu)\ket{1},
		\label{eq:idler_split}
	\end{equation}
	with operators
	\begin{equation}
		\begin{aligned}
			\hat K_{\ell;s}&=\sqrt{\tfrac{\kappa\tau}{4}}\,\big[x_\ell\big(\cosh r\,\hat a+i\nu^*\sinh r\,\hat{\mathbb I}\big)+y_\ell\hat{\mathbb I}\big],\\
			\hat K_{\ell;n}&=\sqrt{\tfrac{\kappa\tau}{4}}\;i\,x_\ell\sinh r\,\hat{\mathbb I}.
		\end{aligned}
		\label{eq:signoise}
	\end{equation}
	The states $\ket{\nu}$ and $\hat D(\nu)\ket{1}$ are orthonormal, so $\hat K_{\ell;s}$ and $\hat K_{\ell;n}$ are the Kraus operators of Eq.~\eqref{eq:coll_map} for a click of detector $\ell$ in which the idler bin is left in $\ket{\nu}$ and in $\hat D(\nu)\ket{1}$, respectively. No other basis state contributes, because $\hat K_{\ell}$ is linear in $\hat v^\dagger$.
	
	The Kraus operators $\hat K_{\ell;s}$ and $\hat K_{\ell;n}$ describe clicks of detector $\ell$ caused by the signal and by the idler noise, respectively. In what follows, we take $\nu$ to be real. The signal operators then contain the amplified signal, $\cosh r\,\hat a$, together with the c-number contribution $i\nu\sinh r$ of the coherent idler, which shifts the Q signal by an amount independent of the state of the readout mode. The noise operators are proportional to the identity, so a click generated by a noise operator leaves the readout mode unchanged and carries no information about it. These clicks originate from the vacuum fluctuations of the idler, which are amplified into the readout bin, and they occur with total probability $\kappa\tau\sinh^2 r$ per collision, independently of the state of the readout mode and of $\nu$.
	
	Because the idler bin is discarded, the two operators contribute to the same outcome. A click of detector $\ell$ occurs with probability $p_{\ell}=\langle\hat K_{\ell;s}^\dagger\hat K_{\ell;s}+\hat K_{\ell;n}^\dagger\hat K_{\ell;n}\rangle$ and updates the state of the readout mode as $\hat\rho\to(\hat K_{\ell;s}\hat\rho\hat K_{\ell;s}^\dagger+\hat K_{\ell;n}\hat\rho\hat K_{\ell;n}^\dagger)/p_{\ell}$. The number of distinguishable outcomes thus remains equal to the number of detectors, whereas the dynamics involves eight Kraus operators. The no-click operator of detector $\ell$ is $\hat{\mathbb I}-\tfrac12(\hat K_{\ell;s}^\dagger\hat K_{\ell;s}+\hat K_{\ell;n}^\dagger\hat K_{\ell;n})$, which ensures completeness to first order in $\tau$, as in Sec.~\ref{sec:homodyne}. For real $\nu$, the eight Kraus operators are
	\begin{equation}
		\begin{aligned}
			\hat K_{Q_1;s} &= \sqrt{\tfrac{\kappa\tau}{4}}\big(\cosh r\,\hat a
			\\&\qquad + (i\nu\sinh r - i\sqrt{2}\beta)\,\hat{\mathbb I}\big), \\
			\hat K_{Q_1;n} &= \sqrt{\tfrac{\kappa\tau}{4}}\,i\sinh r\,\hat{\mathbb I}, \\
			\hat K_{Q_2;s} &= \sqrt{\tfrac{\kappa\tau}{4}}\big((\nu\sinh r
			+ \sqrt{2}\beta)\,\hat{\mathbb I} \\&\qquad - i\cosh r\,\hat a\big), \\
			\hat K_{Q_2;n} &= \sqrt{\tfrac{\kappa\tau}{4}}\sinh r\,\hat{\mathbb I}, \\
			\hat K_{I_1;s} &= \sqrt{\tfrac{\kappa\tau}{4}}\big(-i\cosh r\,\hat a
			\\&\qquad + (\nu\sinh r + i\sqrt{2}\beta)\,\hat{\mathbb I}\big), \\
			\hat K_{I_1;n} &= \sqrt{\tfrac{\kappa\tau}{4}}\sinh r\,\hat{\mathbb I}, \\
			\hat K_{I_2;s} &= \sqrt{\tfrac{\kappa\tau}{4}}\big((-i\nu\sinh r
			- \sqrt{2}\beta)\,\hat{\mathbb I} \\&\qquad - \cosh r\,\hat a\big), \\
			\hat K_{I_2;n} &= -\sqrt{\tfrac{\kappa\tau}{4}}\,i\sinh r\,\hat{\mathbb I}.
			\label{eq:gain_hetero_kraus}
		\end{aligned}
	\end{equation}
	For $\nu=0$, the signal operators reduce to Eq.~\eqref{eq:hetero_kraus} with $\hat a\to\cosh r\,\hat a$, and for $r=0$ they reduce to Eq.~\eqref{eq:hetero_kraus} itself, with the noise operators vanishing.
	
	With $\hat Q$ and $\hat I$ as in Sec.~\ref{sec:heterodyne}, the signals and the noise contributions to the click probabilities are
	\begin{equation}
		\begin{aligned}
			p_Q &= p_{Q_2}-p_{Q_1}=\sqrt{2}\,\kappa\tau\,\beta\,\big[\cosh r\,\langle\hat Q\rangle+\nu\sinh r\big], \\
			p_I &= p_{I_2}-p_{I_1}=\sqrt{2}\,\kappa\tau\,\beta\cosh r\,\langle\hat I\rangle, \\
			p_{\ell;n} &= \big\langle\hat K_{\ell;n}^\dagger\hat K_{\ell;n}\big\rangle=\tfrac{\kappa\tau}{4}\sinh^2 r .
		\end{aligned}
		\label{eq:gain_probs}
	\end{equation}
	 The noise contributions $p_{\ell;n}$ are the minimum added noise allowed for a phase-preserving amplifier with power gain $\cosh^2 r$~\cite{Caves1982,Clerk2010,Caves_2012}. Since a coherent idler has the same fluctuations as the vacuum, the added noise remains minimal for any $\nu$, and the coherent amplitude adds only a deterministic offset to the record and, as shown in Sec.~\ref{sec:backaction}, a drive. Here, the quantum limit is read off directly from the noise Kraus operators.
	
	\subsection{\label{sec:gain_snr}Signal-to-noise ratio}
	
\begin{figure}
		\centering
		\includegraphics{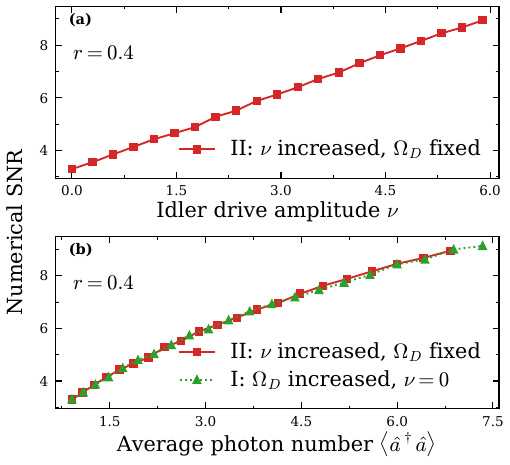}
		\caption{SNR of the gain--heterodyne readout [Fig.~\ref{fig:all-gates}(d)] with a coherent idler, for gain parameter $r=0.4$. (a)~Numerical SNR versus the idler amplitude $\nu$ at fixed readout drive $\Omega_D/2\pi=1$~MHz (protocol~II). The SNR increases with $\nu$. (b)~The same data (red squares, II) plotted against the average intracavity photon number $\langle\hat a^\dagger\hat a\rangle$, together with a vacuum-idler protocol in which $\nu=0$ and $\Omega_D$ is increased (green triangles, I). The two curves coincide, showing that the SNR increase in (a) is set by the photon number, i.e., by the total effective drive $\Omega=\Omega_D+\tfrac{\kappa}{4}\sinh 2r\,\nu$. This increase is due entirely to the additional coherent drive and not to an improvement of the measurement itself. Parameters: $\kappa/2\pi=1$~MHz, $\chi/2\pi=1.5$~MHz, $|\beta|^2=400$, and $t_{\rm int}=1000$~ns.}
		\label{fig:snr_gain}
	\end{figure}

	We now use the Kraus operators of Eq.~\eqref{eq:gain_hetero_kraus} to quantify how the gain affects the dispersive readout of Sec.~\ref{sec:level4}. For a qubit in the basis state $\ket{j}$, $j\in\{e,g\}$, the outcomes of repeated shots form a blob in the IQ plane (Sec.~\ref{sec:assign}), and we quantify the distinguishability of the two blobs by the SNR
	\begin{equation}
		\mathrm{SNR}=\frac{|\mu_e-\mu_g|}{\sqrt{\sigma_e^2+\sigma_g^2}},
		\label{eq:snr_blobs}
	\end{equation}
	where $\mu_j$ is the center of the blob for $\ket{j}$, written as a complex number in the IQ plane, and $\sigma_j^2$ is its total variance, i.e., the sum of the variances of its I and Q components. As derived in Appendix~\ref{app:snr}, in the regime in which the local oscillator dominates the noise of the counts, $\beta^2\gg\langle\hat a^\dagger\hat a\rangle\cosh^2 r$ and $\beta^2\gg(1+\nu^2)\sinh^2 r$, and for an integration time $t_{\rm int}\gg2/(\kappa\cosh^2 r)$, for which the ring-up of the readout mode can be neglected, the SNR is
	\begin{equation}
		\begin{aligned}
			\mathrm{SNR}&\simeq \sqrt{2\kappa t_{\rm int}}\;
			\frac{\cosh r\,|\chi|\,|\Omega|}{\kappa^2\cosh^4 r/4+\chi^2},
			\\
			\Omega&=\Omega_D+\frac{\kappa}{4}\sinh 2r\,\nu .
		\end{aligned}
		\label{eq:snr_main}
	\end{equation}
	The idler amplitude $\nu$ enters only through the effective drive $\Omega$, so that at fixed $r$ and $\Omega_D$, the SNR grows linearly with $\nu$, as confirmed by the numerical simulation in Fig.~\ref{fig:snr_gain}(a). This raises the question of whether a coherent idler improves the measurement itself, which we address next.
	
	\subsection{\label{sec:backaction}Measurement back-action}
	
	Read on its own, Fig.~\ref{fig:snr_gain}(a) suggests that a coherent idler improves the measurement. The master equation shows that this interpretation is misleading. The eight Kraus operators of Eq.~\eqref{eq:gain_hetero_kraus}, four signal and four noise operators, define the jump operators of a Lindblad master equation for the qubit and the readout mode [cf.\ Eq.~\eqref{eq:disp_master}],
	\begin{equation}
		\begin{aligned}
			\hat H &= -\chi\hat\sigma_z\hat a^\dagger\hat a+\Omega_D(\hat a+\hat a^\dagger),\\
			\hat L_j &= \hat K_{\ell;w}/\sqrt{\tau},\quad j=1,\dots,8,\\
			\ell&\in\{Q_1,Q_2,I_1,I_2\},\ w\in\{s,n\}.
		\end{aligned}
		\label{eq:lindblad_setup}
	\end{equation}
	Here, $\hat H$ is the readout Hamiltonian of Eq.~\eqref{eq:full_ham} for a drive at $\omega_d=\omega_r$. The signal and noise operators of each pair feed the same detector and are experimentally indistinguishable, so the measurement record still consists of the clicks of the four detectors. Each jump operator has the form $\hat L=c\,\hat a+d\,\hat{\mathbb I}$ with complex constants $c$ and $d$, for which
	\begin{equation}
		\mathcal D[\hat L]\hat\rho=|c|^2\mathcal D[\hat a]\hat\rho
		+\tfrac12\big(c\,d^*[\hat a,\hat\rho]-c^*d\,[\hat a^\dagger,\hat\rho]\big).
		\label{eq:D_identity}
	\end{equation}
	The noise operators ($c=0$) therefore drop out of the dynamics, and the local-oscillator contributions cancel between the two detectors of each homodyne module, as in Eq.~\eqref{eq:homodyne_step}. Summing the dissipators over all eight operators gives
	\begin{equation}
		\begin{aligned}
			\sum_{j=1}^{8}\mathcal D[\hat L_j]\hat\rho
			={}&-i\left[\tfrac{\kappa}{4}\sinh 2r\,\nu\,(\hat a+\hat a^\dagger),\ \hat\rho\right]\\
			&+\kappa\cosh^2 r\,\mathcal D[\hat a]\hat\rho ,
		\end{aligned}
		\label{eq:dissipator_sum}
	\end{equation}
	i.e., a dissipator rescaled by $\cosh^2 r$ plus a coherent drive proportional to $\sinh 2r$. The complete master equation is therefore
	\begin{equation}
		\begin{aligned}
			\dot{\hat\rho} &= -i[\hat H_{\mathrm{eff}},\hat\rho]
			+\kappa\cosh^2 r\,\mathcal D[\hat a]\hat\rho,\\
			\hat H_{\mathrm{eff}} &= -\chi\hat\sigma_z\hat a^\dagger\hat a
			+\left(\Omega_D+\tfrac{\kappa}{4}\sinh 2r\,\nu\right)(\hat a+\hat a^\dagger).
		\end{aligned}
		\label{eq:master_eq}
	\end{equation}
	
	For $r=0$, Eq.~\eqref{eq:master_eq} reduces to Eq.~\eqref{eq:disp_master} with $\omega_d=\omega_r$. The gain within the coupling thus acts back on the readout mode in two ways. First, the decay rate increases from $\kappa$ to $\kappa\cosh^2 r$, because the readout mode emits into the readout bin through the exchange term of Eq.~\eqref{eq:U_BG}, whose strength at fixed $g$ is enhanced by $\cosh r$. Second, a coherent idler adds the readout drive $\tfrac{\kappa}{4}\sinh 2r\,\nu\,(\hat a+\hat a^\dagger)$, which adds to $\Omega_D$ and yields exactly the effective drive $\Omega$ of Eq.~\eqref{eq:snr_main}. Physically, the two-mode-squeezing term converts the coherent idler into a phase-conjugated coherent field in the readout bin, which the exchange term feeds back into the readout mode within the same collision. Like the dissipator in Eq.~\eqref{eq:coll_step}, this second-order process contributes at first order in $\tau$, and unlike the local-oscillator terms, it does not cancel between the detectors. In this model, the vacuum fluctuations of the idler, in contrast, do not heat the readout mode; they enter only the measurement record, through the noise operators. The usual description, in which an amplifier enters the SME only through a detection efficiency or added noise~\cite{Gambetta2008}, presupposes an amplifier isolated from the readout mode and therefore misses both effects. Here, they follow directly from composing the gates and summing the dissipators.
	
	
	If the increase of the SNR with $\nu$ originates from the drive, as shown by Eq.~\eqref{eq:snr_main}, the same increase should be obtained by raising the photon number through $\Omega_D$ instead of $\nu$. We test this by comparing two protocols that reach the same intracavity photon number $\langle\hat a^\dagger\hat a\rangle$:
	\begin{itemize}
		\item[(I)] a vacuum idler ($\nu=0$) with increased $\Omega_D$;
		\item[(II)] a coherent idler with increased $\nu$ and fixed $\Omega_D$.
	\end{itemize}
	Figure~\ref{fig:snr_gain}(b) shows that the two curves coincide. The additional $\nu$-dependent offset $\propto\nu\sinh r$ in the IQ plane is the same for both qubit states and cancels in the distance between the two blobs, so $\nu$ affects the SNR only by raising the photon number through $\Omega$. Appendix~\ref{app:snr} confirms this analytically. Protocols I and II have identical mean blob separations when $\Omega_D'=\Omega_D+\tfrac{\kappa}{4}\sinh 2r\,\nu$, and their SNRs differ only through the shot noise of the offset, which gives a relative correction suppressed by $1/\beta^2$. The increase of the SNR with $\nu$ is therefore due entirely to the additional coherent drive and not to an improvement of the measurement itself. A correct accounting of the SNR requires this back-action, and it shows that a coherent idler offers no advantage over simply increasing $\Omega_D$.
	
	Within this model, Eq.~\eqref{eq:master_eq} also allows the gain parameters to be calibrated from measured quantities. The gain parameter follows from the ratio of the measured linewidth $\kappa_{\rm meas}=\kappa\cosh^2 r$ to the bare linewidth $\kappa$, i.e., the linewidth at $r=0$ for the same coupling $g$, and the idler amplitude follows from the measured effective drive $\Omega_{\rm meas}$, obtained, e.g., from the steady-state photon number $\Omega_{\rm meas}^2/(\kappa_{\rm meas}^2/4+\chi^2)$ (Appendix~\ref{app:snr}),
	\begin{equation}
		r=\cosh^{-1}\sqrt{\frac{\kappa_{\rm meas}}{\kappa}},
		\qquad
		\nu=\frac{4(\Omega_{\rm meas}-\Omega_D)}{\kappa\sinh 2r} .
		\label{eq:calibration}
	\end{equation}
	
	Because the gain acts within the coupling, it modifies the dynamics of the readout mode itself: the readout mode decays at the rate $\kappa\cosh^2 r$ and, for a coherent idler, is driven, while each emitted photon still yields a single detected photon. This differs from the conventional configuration of superconducting-qubit readout, in which circulators isolate the readout resonator from an amplifier placed in its output line~\cite{Clerk2010,Abdo2013PRX,Krantz2019,Blais2021}. There, the decay rate remains $\kappa$, and only the measurement record is amplified. Describing that configuration requires the amplifier of finite bandwidth discussed at the beginning of this section.
	
	\section{\label{sec:level6}Summary and outlook}
	
	We have developed a modular linear-optics collision model for the readout of propagating fields. Each element of the detection chain is represented by a gate unitary and each photodetector by a click operator, and the effective Kraus operators on the system alone follow from composing these gates and tracing out the ancillas in their known states. The decay rate of the system enters only through the mixing angle of the collision, and since at most one photon reaches each detector per collision, the click operators of the photodetectors are simply the annihilation operators of the detected modes. The construction requires no stochastic calculus, retains individual photocounts, and reproduces the standard master equation and measurement operators in the continuous-time limit. Because every element enters as an independent gate, a more complex scheme is built by inserting or replacing gates and reusing the Kraus operators of a simpler scheme as a module. As a first example, we derived the homodyne Kraus operators, which coincide with the measurement operators of the standard description, and used them to construct the heterodyne Kraus operators, which reproduce the reduction of the signal per quadrature by a factor of $\sqrt{2}$ relative to homodyne detection, the familiar 3-dB penalty~\cite{YamamotoHaus1986,CavesDrummond1994,Clerk2010}.
	
	Applied to dispersive qubit readout, the heterodyne module reveals two effects that a description in terms of two fixed Gaussian blobs does not capture. First, treating the qubit and the readout mode jointly reveals a fundamental assignment error that depends on the initial qubit state and is present at any finite integration time even in an ideal readout chain. It is largest near the $\hat\sigma_z$ eigenstates, where it equals the overlap of the two blobs, and smallest for $|+\rangle$, because the record that determines the assignment also drives the collapse of a superposition. Second, because the model retains the discreteness of the photocounts, the IQ distribution crosses over from Skellam to Gaussian as the local-oscillator amplitude grows, with the crossover at an amplitude of order unity identified by the KL divergence, a regime beyond the diffusive SME.
	
	As a second example, we included phase-preserving gain by replacing only the collision gate with a three-mode gate that combines a beam-splitter coupling with two-mode squeezing, while reusing the heterodyne module unchanged. This keeps the number of photons per collision of order $\tau$ and yields eight Kraus operators, four signal and four noise operators, for the four detectors of the heterodyne module. The noise operators are proportional to the identity and carry the minimum added noise of a phase-preserving amplifier. Because the gain acts within the coupling, it acts back on the readout mode: the decay rate increases to $\kappa\cosh^2 r$, and a coherent idler adds the drive $\tfrac{\kappa}{4}\sinh 2r\,\nu\,(\hat a+\hat a^\dagger)$. This drive explains why the SNR grows with $\nu$. It raises the intracavity photon number exactly as an increased $\Omega_D$ would, and a vacuum-idler protocol with matched photon number gives the same SNR up to a small correction that vanishes for a strong local oscillator. The improvement is therefore due entirely to the additional drive, not to an improvement of the measurement itself, and a coherent idler offers no advantage over simply increasing $\Omega_D$. A description that includes the gain only through an efficiency or an added-noise term, which presupposes an amplifier isolated from the readout mode, would miss this back-action.
	
	Several extensions follow naturally. Describing an amplifier placed after the readout resonator, as in standard readout chains with a circulator, requires an amplifier of finite bandwidth, included as an additional cascaded mode in the collision model. We leave this extension to future work. The same consideration applies to phase-sensitive amplification before detection, which would extend the model to the full family of diffusive unravelings~\cite{WisemanDiosi2001,ChiaWiseman2011}. Further gates could model photon loss, finite detector efficiency, or filtering, and the framework could be applied to multiplexed readout and to feedback based on single trajectories. Because the model provides the conditional state of the qubit alongside each measurement record, it could also be used to optimize the filtering of the record. Comparing the simulated IQ statistics and assignment errors with experimental readout data would test the model on real detection chains and enable their optimization. Finally, since the approach is modular and general, it can be applied to a wide range of settings, including those relevant to quantum communication and quantum sensing.

	{\it Acknowledgements---} AP gratefully thanks the FRIENDSHIP 2.0 program of Japan International Cooperation Agency (JICA) and IIT Hyderabad Seed Grant for support.

	\appendix

\section{Derivation of the SNR for the gain--heterodyne readout}
\label{app:snr}

Throughout this appendix, $\beta$ and $\nu$ are real. Each operator in Eq.~\eqref{eq:gain_hetero_kraus} has the form $\hat K_j=\sqrt{\kappa\tau/4}\,(c_j\hat a+d_j\hat{\mathbb I})$, where $j$ labels the eight channels $\ell;w$ of Eq.~\eqref{eq:lindblad_setup}, with the coefficients listed in Table~\ref{tab:coeffs}.

\begin{table}[b]
	\caption{\label{tab:coeffs}Coefficients $c_j$ and $d_j$ of the gain--heterodyne Kraus operators $\hat K_j=\sqrt{\kappa\tau/4}\,(c_j\hat a+d_j\hat{\mathbb I})$ of Eq.~\eqref{eq:gain_hetero_kraus}.}
	\begin{ruledtabular}
		\begin{tabular}{lcc}
			Channel & $c_j$ & $d_j$ \\
			\colrule
			$Q_1;s$ & $\cosh r$   & $i(\nu\sinh r-\sqrt2\beta)$ \\
			$Q_2;s$ & $-i\cosh r$ & $\nu\sinh r+\sqrt2\beta$ \\
			$I_1;s$ & $-i\cosh r$ & $\nu\sinh r+i\sqrt2\beta$ \\
			$I_2;s$ & $-\cosh r$  & $-(\sqrt2\beta+i\nu\sinh r)$ \\
			$\ell;n$, all $\ell$ & $0$ & $|d_j|=\sinh r$ \\
		\end{tabular}
	\end{ruledtabular}
\end{table}

\subsection{Steady state}

With the coefficients of Table~\ref{tab:coeffs}, $\sum_j|c_j|^2=4\cosh^2 r$ and $\sum_j c_jd_j^*=-4i\nu\cosh r\sinh r$, where the $\beta$-dependent terms cancel between the four channels. By Eq.~\eqref{eq:D_identity}, the eight operators therefore give the decay rate $\kappa\cosh^2 r$ and the additional drive $\tfrac{\kappa}{4}\sinh 2r\,\nu$, which adds to $\Omega_D$ to give the effective drive $\Omega$ of Eq.~\eqref{eq:snr_main} [cf.\ Eq.~\eqref{eq:master_eq}]. The field amplitude of the readout mode thus decays at the rate
\begin{equation}
	\gamma\equiv\frac{\kappa}{2}\cosh^2 r ,
	\label{eq:k_def}
\end{equation}
and we define
\begin{equation}
	M\equiv \gamma^2+\chi^2 .
	\label{eq:M_def}
\end{equation}
For the qubit in its excited state $\ket{e}$ or its ground state $\ket{g}$, the readout mode is detuned from the drive by $\mp\chi$, and its steady state is the coherent state of Eq.~\eqref{eq:alpha_sym} with $\kappa/2\to\gamma$ and $\Omega_D\to\Omega$,
\begin{equation}
	\begin{aligned}
		\alpha_{e/g}&=\frac{\Omega(\pm\chi-i\gamma)}{M},\qquad
		\langle\hat a^\dagger\hat a\rangle=\frac{\Omega^2}{M},\\
		\alpha_e-\alpha_g&=\frac{2\chi\Omega}{M},\qquad
		\alpha_e+\alpha_g=\frac{-2i\gamma\,\Omega}{M}.
	\end{aligned}
	\label{eq:ss}
\end{equation}

\subsection{Mean signal and noise}

The mean count of channel $j$ is $N_j=\tfrac{\kappa\tau}{4}\big(|c_j|^2\langle\hat a^\dagger\hat a\rangle+2\,\mathrm{Re}(c_jd_j^*\alpha)+|d_j|^2\big)$. With $u_I=N_{I_2}-N_{I_1}$ and $u_Q=N_{Q_2}-N_{Q_1}$ (the noise channels cancel in each difference), the mean point in the IQ plane per collision is
\begin{equation}
	u=u_I+iu_Q=\sqrt2\,\kappa\tau\,\beta\,\big(\alpha\cosh r+i\nu\sinh r\big).
	\label{eq:u}
\end{equation}
This is $p_I+ip_Q$ of Eq.~\eqref{eq:gain_probs}. The photon number drops out, and $i\nu\sinh r$ is a state-independent offset that cancels between $\ket{e}$ and $\ket{g}$. Hence, using Eq.~\eqref{eq:ss},
\begin{equation}
	\begin{aligned}
		|u_e-u_g|&=\sqrt2\,\kappa\tau\,\beta\cosh r\,|\alpha_e-\alpha_g|\\
		&=\frac{2\sqrt2\,\kappa\tau\,\beta\cosh r\,|\chi|\,|\Omega|}{M}.
	\end{aligned}
	\label{eq:dist}
\end{equation}
The counts are Poissonian, so the variance equals the mean. Summing all eight channels, with $\sum_j|d_j|^2=8\beta^2+4\nu^2\sinh^2 r$ for the four signal channels and four noise channels each contributing $\tfrac{\kappa\tau}{4}\sinh^2 r$, gives the total variance of the IQ point per collision, i.e., the sum of the variances of its I and Q components, for the qubit in $\ket{e}$ or $\ket{g}$,
\begin{equation}
	\begin{aligned}
		\tilde\sigma_{e/g}^2&=\sum_{j=1}^8 N_j\\
		&=\kappa\tau\Big[\langle\hat a^\dagger\hat a\rangle\cosh^2 r
		+2\nu\cosh r\sinh r\,\mathrm{Im}\,\alpha\\
		&\quad+2\beta^2+\nu^2\sinh^2 r+\sinh^2 r\Big].
	\end{aligned}
\end{equation}
Adding these variances for $\ket{e}$ and $\ket{g}$ and using Eq.~\eqref{eq:ss}, we obtain
\begin{equation}
	\begin{aligned}
		\tilde\sigma^2&=\tilde\sigma_e^2+\tilde\sigma_g^2=2\kappa\tau\,W,\\
		W&=\frac{\Omega^2\cosh^2 r-2\gamma\nu\Omega\cosh r\sinh r}{M}\\
		&\quad+\nu^2\sinh^2 r+2\beta^2+\sinh^2 r .
	\end{aligned}
	\label{eq:W}
\end{equation}

\subsection{SNR}

For an integration time $t_{\rm int}\gg1/\gamma$, the ring-up of the readout mode can be neglected (cf.\ Sec.~\ref{sec:disp_model}). The counts of different collisions are then independent and identically distributed, so the blob centers and total variances of Eq.~\eqref{eq:snr_blobs} are $\mu_{e/g}=(t_{\rm int}/\tau)\,u_{e/g}$ and $\sigma_{e/g}^2=(t_{\rm int}/\tau)\,\tilde\sigma_{e/g}^2$. Inserting Eqs.~\eqref{eq:dist} and \eqref{eq:W} into Eq.~\eqref{eq:snr_blobs} therefore amounts to replacing $\tau\to t_{\rm int}$ in $|u_e-u_g|/\tilde\sigma$, and we obtain
\begin{equation}
	\mathrm{SNR}
	=\frac{2\sqrt{\kappa t_{\rm int}}\;\beta\cosh r\,|\chi|\,|\Omega|}{M\sqrt{W}} .
	\label{eq:snr_full}
\end{equation}
When the local oscillator dominates the noise, $W=2\beta^2(1+\zeta)$ with $\zeta=\big[\Omega^2\cosh^2 r/M-2\gamma\nu\Omega\cosh r\sinh r/M+\nu^2\sinh^2 r+\sinh^2 r\big]/2\beta^2\ll1$, which requires $\beta^2\gg\langle\hat a^\dagger\hat a\rangle\cosh^2 r$ and $\beta^2\gg(1+\nu^2)\sinh^2 r$. Then $\beta$ cancels in Eq.~\eqref{eq:snr_full},
\begin{equation}
	\begin{aligned}
		\mathrm{SNR}&\simeq\sqrt{2\kappa t_{\rm int}}\;\frac{\cosh r\,|\chi|\,|\Omega|}{M}\\
		&=\sqrt{2\kappa t_{\rm int}}\;\cosh r\,|\chi|\sqrt{\frac{\langle\hat a^\dagger\hat a\rangle}{M}},
	\end{aligned}
	\label{eq:snr_lo}
\end{equation}
which is Eq.~\eqref{eq:snr_main}. In this regime, the SNR is proportional to the total drive $\Omega$ and grows linearly with $\nu$ at fixed $\Omega_D$ through $\Omega=\Omega_D+\tfrac{\kappa}{4}\sinh2r\,\nu$.

\subsection{Equivalence of the two protocols}

Consider protocol (I), with $\nu=0$ and drive $\Omega_D'$, and protocol (II), with $\nu\neq0$ and drive $\Omega_D$. Both have the same $\Omega$, and hence the same photon number $\langle\hat a^\dagger\hat a\rangle=\Omega^2/M$, when
\begin{equation}
	\Omega_D'=\Omega_D+\frac{\kappa}{4}\sinh 2r\,\nu .
\end{equation}
The mean distance of Eq.~\eqref{eq:dist} is then identical, since $\nu$ enters only through $\Omega$. The only remaining difference is in the noise, $W_{\rm II}-W_{\rm I}=\nu^2\sinh^2 r-2\gamma\nu\Omega\cosh r\sinh r/M$, which arises from the shot noise of the offset $i\nu\sinh r$ in Eq.~\eqref{eq:u} and is small compared with $W\simeq2\beta^2$. Hence, to leading order in $1/\beta^2$, $\mathrm{SNR}_{\rm II}/\mathrm{SNR}_{\rm I}=1-(W_{\rm II}-W_{\rm I})/(4\beta^2)$, in agreement with the coincidence of the two curves in Fig.~\ref{fig:snr_gain}(b). The small residual offset at the largest photon numbers is of this order. The increase of the SNR with $\nu$ is thus entirely due to the back-action drive $\tfrac{\kappa}{4}\sinh2r\,\nu(\hat a+\hat a^\dagger)$ and not to an improvement of the measurement itself.

	\bibliography{refv3}
	
\end{document}